\documentclass[12pt]{article}
\usepackage[latin9]{inputenc}
\usepackage{amsmath}
\usepackage{amssymb}

\makeatletter

\usepackage{color}

\usepackage{a4}
\usepackage{verbatim}\numberwithin{equation}{section}

\newtheorem{theorem}{Theorem}[section]
\newtheorem{lemma}[theorem]{Lemma}\newtheorem{remark}[theorem]{Remark}\newtheorem{corollary}[theorem]{Corollary}\newtheorem{proposition}[theorem]{Proposition}\newtheorem{definition}[theorem]{Definition}\newtheorem{ex}{Example}[section]
\newenvironment{example}{\begin{ex}\rm}{ \hfill $\Diamond$ \end{ex}
        \vskip4pt}
\newtheorem{ass}{Assumption}[section]

\numberwithin{equation}{section}

\makeatother

\begin{document}
\global\long\def\dx#1{\frac{\partial}{\partial#1}}%
\global\long\def\dh#1{\mathop{#1}\limits _{h}}%
\global\long\def\dhp#1{\mathop{#1}\limits _{+h}}%
\global\long\def\dhm#1{ \mathop{#1}\limits _{-h}}%
\global\long\def\dphh#1{ \mathop{#1}\limits _{h \bar{h}}}%
\global\long\def\da#1{ \mathop{#1}\limits _{+\tau}}%
\global\long\def\db#1{ \mathop{#1}\limits _{-\tau}}%
\global\long\def\dc#1{ \mathop{#1}\limits _{\pm\tau}}%
\global\long\def\dd#1{ \mathop{#1}\limits _{+h}}%
\global\long\def\df#1{ \mathop{#1}\limits _{-h}}%
\global\long\def\dpm#1{ \mathop{#1}\limits _{\pm h}}%
\global\long\def\dg#1{ \mathop{#1}\limits _{\pm h}}%
\global\long\def\dh#1{ \mathop{#1}\limits _{h}}%
\global\long\def\sso#1{\ensuremath{\mathfrak{#1}}}%
\global\long\def\ddt{\frac{\partial}{\partial t}}%
\global\long\def\ddx{\frac{\partial}{\partial x}}%
\global\long\def\ddy{\frac{\partial}{\partial y}}%
\global\long\def\ddyy{\frac{\partial}{\partial y'}}%

\newcommand{\DD}{D}     
\newcommand{\Dt}{D_t}
\newcommand{\Dtm}{D_{t^-}}
\newcommand{\Dtp}{D_{t^+}}

\newcommand{\deltau}{{\delta  \over \delta  q} _{(E)}}
\newcommand{\deltauL}{\left. {\delta L \over \delta  q} \right |  _{(E)}}
\newcommand{\deltat}{\delta  \over \delta t}
\newcommand{\deltatL}{\delta L \over \delta t}

\begin{center}
\textbf{\Large  Delay ordinary differential equations: \\
 from Lagrangian approach  to Hamiltonian approach}
\end{center}




\bigskip

\begin{center}
{\large Vladimir Dorodnitsyn}$^{a}$, 
{\large Roman Kozlov}$^{b}$, 
{\large Sergey Meleshko}$^{c}$
\end{center}

\bigskip{}

\vspace*{10mm}

\noindent $^{a}$ Keldysh Institute of Applied Mathematics, Russian
Academy of Science, \\
 Miusskaya Pl.~4, Moscow, 125047, Russia; \\
 {e-mail: Dorodnitsyn@Keldysh.ru, dorod2007@gmail.com} \\
$^{b}$ Department of Business and Management Science, Norwegian School of Economics, \\
Helleveien 30, 5045, Bergen, Norway; \\
 {e-mail: Roman.Kozlov@nhh.no} \\
$^{c}$ School of Mathematics, Institute of Science, Suranaree University of Technology, \\
30000, Thailand; \\
 {e-mail: sergey@math.sut.ac.th} \\

\begin{center}
\textbf{Abstract}
\end{center}

\begin{quotation}
The paper suggests a Hamiltonian formulation for delay ordinary
differential equations (DODEs). Such equations are related to DODEs
with a Lagrangian formulation via a delay analog of the Legendre
transformation. The Hamiltonian delay operator identity is
established. It states the relationship for the invariance of a
delay Hamiltonian functional, appropriate delay variational
equations, and their conserved quantities. The identity is used to
formulate a Noether-type theorem, which provides first integrals
for Hamiltonian DODEs with symmetries. 
The relationship between the invariance of the delay Hamiltonian functional 
and the invariance of the delay variational equations is also examined. 
Several examples illustrate the theoretical results.
\end{quotation}

\bigskip
\bigskip



\eject

\section{Introduction}

Applications of Lie groups of transformations to differential equations,
as well as to {integro}differential and other {non}local equations,
originate from the classical work of Sophus Lie~\cite{bk:Lie[1888], bk:Lie1924}.
Lie group symmetries of differential equations can be utilized to find exact analytic solutions
(solutions invariant under certain groups),
transform solutions into other solutions,
and classify equations into equivalence classes
\cite{bk:Ovsiannikov1978,
bk:Ibragimov[1983], bk:Olver[1986], bk:Gaeta1994, bk:HandbookLie,
bk:BlumanAnco2002}.
Since the seminal work of E. Noether~\cite{Noether1918}, symmetry groups have provided a foundation for deriving first integrals and conservation laws for differential equations with a variational formulation.
The extension of this connection to differential equations without a Lagrangian or Hamiltonian formulation
was developed  in~\cite{bk:AncoBluman1997, bk:BlumanAnco2002}.

Lie groups of transformations have been extended to difference, discrete, and differential-difference equations
\cite{Dorodnitsyn1991, LeviWinternitz1991, QuispelCapelSahadevan,
Dorodnitsyn1993a, Dorodnitsyn1993, DorodnitsynKozlovWinternitz2000,
DorKozWin2004,
 LeviWinternitz2005,
bk:Dorodnitsyn[2011], Winternitz2011,
bk:Hydon2014, bk:DKapKozWin, bk:Zhang}.
Symmetries have been employed to construct numerical schemes
that preserve qualitative properties of the  differential equations~\cite{bk:ChevDK, bk:DKap1, bk:DKap2,  bk:DKapMel}.

The present paper is part of a research work to extend the
application of group analysis to delay ordinary differential
equations (DODEs)~\cite{bk:Elsgolts[1955], Driver1977,  Erneux,
Polyanin2023}. In previous
publications~\cite{bk:DorodnitsynKozlovMeleshkoWinternitz[2018a],
bk:DorodnitsynKozlovMeleshkoWinternitz[2018b]}, the Lie group
classification of first-order delay ordinary differential equations
was presented. The group classification of second-order delay
ordinary differential equations was detailed
in~\cite{bk:DorodnitsynKozlovMeleshkoWinternitz2021}. In the recent
article~\cite{bk:DorodnitsynKozlovMeleshko[2023]}, 
the Lagrangian formalism for delay ODEs was developed, and a Noether-type theorem
for invariant variational problems was established.



The present paper is devoted to the Hamiltonian approach to DODEs.
In this article, which is based on the results of~\cite{bk:DorodnitsynKozlovMeleshko[2023]},
we explore the connection between the Lagrangian and Hamiltonian descriptions
for delay ordinary differential equations.
We introduce delay canonical Hamiltonian equations,
which are compatible with a delay analog of the Euler-Lagrange equation.
The compatibility is based on a delay version of the Legendre transformation.
We establish an analog of the Noether operator identity for the Hamiltonian approach.
This identity relates the invariance of a delay Hamiltonian functional
to corresponding delay variational equations and their conserved quantities.
Using this identity, we formulate a Noether-type theorem
that provides first integrals for Hamiltonian DODEs.
If there are sufficiently many first integrals,
they can be used for the recursive presentation of the solutions. 
The relationship between the invariance of the delay Hamiltonian functional 
and the invariance of the delay variational equations is also examined.  
Several examples are presented to illustrate the theoretical results.

The paper is organized as follows.
Section~\ref{section_Variational_approach} reminds the variational framework to Hamiltonian ODEs.
A variational framework to delay ordinary differential equations is considered in section~\ref{section_delay_ODEs}.
A Hamiltonian formalism for DODEs is developed in section~\ref{section_Hamiltonian_formalism}.
Section~\ref{section_symmetries} introduces symmetries
for delay variational equations in the Hamiltonian framework 
and states the infinitesimal criteria for invariance of delay Hamiltonian functionals.
Two types of first integrals (differential and difference) of delay Hamiltonian equations
are defined in section~\ref{section_first_integrals}.
Section~\ref{section_Noether} provides the Hamiltonian version of the Noether  operator identity
and the Noether theorem.
The relationship between the invariance of delay Hamiltonian functionals
and the invariance of delay variational equations is given in  section~\ref{section_Invariance}.
Section~\ref{Section_Examples} presents examples that illustrate theoretical results.
Final section~\ref{section_conclusion}  contains concluding remarks.
Complementary results are separated into Appendices.


\section{Hamiltonian ODEs}
\label{section_Variational_approach}

It is well known~\cite{Gelfand,  Abraham, Goldstein, Arnold, Marsden}
that variations of the action functional
\begin{equation}   \label{functional_L}
{\cal L}
=
\int_{a}^{b}
L (t,{q},\dot{q})  \ dt,
\end{equation}
where   $L (t,{q},\dot{q})  $ is a Lagrangian function,
lead to the Euler-Lagrange equation
\begin{equation}   \label{Euler_Lagrange}
{ \delta {L}   \over \delta q}
=
\frac{\partial L}{\partial q}
- D   \left( \frac{\partial L }{\partial  \dot{q}}  \right)
= 0,
\end{equation}
where
\begin{equation*}
 D
=\frac{\partial}{\partial t}
+\dot{q} \frac{\partial}{\partial q}
+\ddot{q} \frac{\partial}{\partial \dot{q} }
+  \cdots
 \end{equation*}
is the total differentiation operator.
We consider a one-dimensional case  $  q \in   \mathbb{R}$ for simplicity.

Lie point symmetries of the  Euler-Lagrange equations~(\ref{Euler_Lagrange})
are given by the generators of the form
\begin{equation}  \label{symmetry_L}
X
= \xi(t,{q})\frac{\partial}{\partial t}
+ \eta(t,{q})\frac{\partial}{\partial q},
\end{equation}
which are prolonged to the derivatives 
according to the standard prolongation formulas~\cite{bk:Ovsiannikov1978, bk:Ibragimov[1983], bk:Olver[1986]}.

In this section, we recall  how one can obtain Hamiltonian ODEs
using  a variational principle and how one can use this variational framework to formulate
the Noether theorem~\cite{bk:DorodnitsynKozlov}.
In the following sections,
we extend these results to delay ordinary differential equations.

\subsection{Variational framework for Hamiltonian ODEs}

 We shortly overview the results of~\cite{bk:DorodnitsynKozlov}
(see also~\cite{bk:DorodnitsynKozlov2009, bk:DorodnitsynKozlov2010})
for scalar dependent functions ${q (t) }$  and  $ {p (t) }$.
Later,  we will follow a similar way to construct Hamiltonian formalism for DODEs.

We consider the canonical Hamiltonian equations~\cite{Goldstein, Arnold}
\begin{equation}    \label{canonical}
\dot{q}=\frac{\partial H}{\partial{p}},
\qquad
\dot{p}=-\frac{\partial H}{\partial{q}},
\end{equation}
for some Hamiltonian function $H=H(t,{q},{p})$.
These equations can be obtained by the variational principle in the phase space $({q},{p})$ from
the action functional
\begin{equation}   \label{functional_H}
{\cal H}
=
\int_{a}^{b} \left(  p \dot{q} - H(t,{q},{p}) \right) d t
\end{equation}
(see, for example,~\cite{Gelfand}).

Let us note that the canonical Hamiltonian equations~(\ref{canonical})
can be obtained by the action of the variational operators
\begin{equation*}   
\frac{\delta}{\delta p}
=
\frac{\partial}{\partial p}
- D\frac{\partial}{\partial\dot{p}},
\qquad
\frac{\delta}{\delta q}
=
\frac{\partial}{\partial q}
-D \frac{\partial}{\partial\dot{q}},
\end{equation*}
where $D$ is the operator of total differentiation with respect to time
\begin{equation*}  
 D
=\frac{\partial}{\partial t}
+\dot{q}\frac{\partial}{\partial q}
+\dot{p}\frac{\partial}{\partial p}+  \cdots,
 \end{equation*}
on the integrand of the functional~(\ref{functional_H}),
namely on the function
\begin{equation}    \label{integrand}
\tilde{H}  =  p \dot{q} - H(t, q,{p}).
\end{equation}
In detail, we obtain
\begin{equation}   \label{varoperator2}
\frac{\delta  \tilde{H}   }{\delta p}
=
\dot{q}
-  \frac{\partial H} {\partial  p }
=  0,
\qquad
\frac{\delta  \tilde{H}   }{\delta q}
=
- \dot{p}
-  \frac{\partial H} {\partial  q }
= 0.
\end{equation}

\begin{remark}   \label{the_third_equation}
We can also consider   the variation with respect to the independent  variable $t$
\begin{equation}     \label{varoperator3}
{\delta  \tilde{H}   \over \delta t }
=
D(H) - \frac{\partial H}{\partial t}
=0.
\end{equation}
This equation holds on the solutions of the canonical Hamiltonian equations~(\ref{canonical}):
\begin{equation*}  
 \left.D(H)\right|_{\dot{q}=H_{p},\
\dot{p}=-H_{q}}
 =\left[
\frac{\partial H}{\partial t}
+\dot{q}\frac{\partial H}{\partial q}
 +\dot{p}\frac{\partial H}{\partial p}
\right]_{\dot{q}=H_{p}, \ \dot{p}=-H_{q}}
=\frac{\partial H}{\partial t}.
 \end{equation*}
The equation~(\ref{varoperator3}) is provided by the action of the variational operator
\begin{equation}       \label{varoperator3_b}
\frac{\delta}{\delta t}
=
\frac{\partial}{\partial t}
+  D  \left(     \dot{q}   \frac{\partial}{\partial\dot{q}  }
  +  \dot{p}   \frac{\partial}{\partial\dot{p}  }      \right)
- D
\end{equation}
on the function~(\ref{integrand}).
\end{remark}

The relationship of the Hamiltonian function  $ H(t,{q},{p})$
with  the Lagrangian function $L(t,{q},{\dot{q}})$
is given by the Legendre transformation
\begin{equation}  \label{Legendre}
H (t,{q},{p})
= p \dot{q}
- L (t,{q},{\dot{q}}),
\end{equation}
where  we should substitute $ \dot{q} $  expressed from
\begin{equation}  \label{to_find_dot_q}
{p}=\frac{\partial L}{\partial\dot{q}}  (t,{q},{\dot{q}})
\end{equation}
in the left-hand side~\cite{Gelfand}.
Equation~(\ref{to_find_dot_q}) can be resolved for  $ \dot{q} $
if  $ \frac{\partial^2 L}{\partial  \dot{q} ^2}   \neq 0 $. 
In detail, the introduction of the canonical Hamiltonian equations 
via the Legendre transformation is reviewed in Appendix~\ref{Comment_derivation}.

The Legendre relation~(\ref{Legendre}) makes it possible
to establish equivalence of Euler--Lagrange equation~(\ref{Euler_Lagrange})
and canonical Hamiltonian equations~(\ref{canonical})
(see, for example,~\cite{Arnold}).
In addition to the relation~(\ref{to_find_dot_q}),  there hold
\begin{equation}
\dot{q}  ={\partial H \over \partial p},
\qquad
{\partial H \over \partial q} = -   {\partial L \over \partial q},
 \qquad
{\partial H \over \partial t}  = -   {\partial L \over \partial t}.
\end{equation}


Lie point symmetries of canonical Hamiltonian equations~(\ref{canonical}) are given
by the generators of the form
\begin{equation}  \label{symmetry_H}
X
= \xi(t,{q},{p})\frac{\partial}{\partial t}
+ \eta(t,{q},{p})\frac{\partial}{\partial q}
+ \nu(t,{q},{p})\frac{\partial}{\partial p}, 
\end{equation}
which are prolonged to the derivatives 
according to the standard prolongation formulas~\cite{bk:Ovsiannikov1978, bk:Ibragimov[1983], bk:Olver[1986]}.

\begin{remark}
It should be noticed that the Legendre transformation~(\ref{Legendre}) is not a point transformation.
Hence, there is no one-to-one correspondence
between  Lie point symmetries of the Euler-Lagrange  equation~(\ref{Euler_Lagrange}),
which are given by the generators of the form~(\ref{symmetry_L}),
and Lie point symmetries of the canonical Hamiltonian equations,
which are presented by the generators of the form~(\ref{symmetry_H}).
\end{remark}

\subsection{Invariance of Hamiltonian functional}
\label{invariance_section}

We consider group transformations in the space $(t,{q},{p})$
which are generated by operators~(\ref{symmetry_H}).
It is shown in~\cite{bk:DorodnitsynKozlov}
that invariance of Hamiltonian functional~(\ref{functional_H})
is equivalent to the invariance of the Hamiltonian elementary action
\begin{equation}  \label{action}
pd{q}-H (t,q,p) d t.
\end{equation}
The Hamiltonian elementary action can be considered as
an analog of the Lagrangian elementary action $ L dt$
corresponding to the functional~(\ref{functional_L}).
A Hamiltonian function $H$ is called invariant with respect to operator~(\ref{symmetry_H})
if the elementary action~(\ref{action}) is an invariant of the group transformation generated by this operator.
Invariance of the elementary action with respect to a group transformation generated
by operator~(\ref{symmetry_H}) prolonged for differentials $dt$ and $dq$
\begin{equation*}
X
= \xi \frac{\partial}{\partial t}
+ \eta \frac{\partial}{\partial q}
+ \nu \frac{\partial}{\partial p}
+  D ( \xi ) d t  \frac{\partial}{\partial d t}
+  D ( \eta )  d t \frac{\partial}{\partial d q}
\end{equation*}
provides the following  infinitesimal invariance criteria~\cite{bk:DorodnitsynKozlov}.

\begin{theorem} \label{Haminvariance}
A Hamiltonian is invariant with respect to a group transformation generated by the operator~(\ref{symmetry_H})
if and only if the following condition holds
\begin{equation}   \label{invar}
X ( \tilde{H} ) + \tilde{H}  D( \xi )
= \nu\dot{q}+pD({\eta})-X(H)-HD(\xi)
=0.
\end{equation}
\end{theorem}

\begin{remark} From the relation
\begin{equation*}   
L(t,{q},{\dot{q}})dt=pd{q}-H(t,{q},{p})dt,
\end{equation*}
it follows that if a Lagrangian is invariant
(i.e., the Lagrangian functional~(\ref{functional_L}) is invariant)
with respect to a group of Lie point transformations with generator~(\ref{symmetry_L}),
the Hamiltonian is also invariant with respect to the same group of point transformations, 
extended to $p$ introduced by~(\ref{to_find_dot_q}).  
The converse statement is false.
\end{remark}

\subsection{The Hamiltonian identity and Noether--type theorem}
\label{identity}

Now, we can relate the conservation properties of the canonical Hamiltonian
equations to the invariance of the Hamiltonian function following~\cite{bk:DorodnitsynKozlov}.

\begin{lemma}   \label{lemmaidentity}
{\bf (The Hamiltonian identity)} The identity
\begin{multline}   \label{ident}
 \nu\dot{q}  +  p D(\eta)  -  X(H)  -  HD(\xi)
\\
\equiv
\xi\left(D(H)-\frac{\partial H}{\partial t}\right)
-    \eta\left(\dot{p}+\frac{\partial H}{\partial q}\right)
+\nu\left(\dot{q}-\frac{\partial H}{\partial p}\right)
+ D\left[p\eta-\xi H\right]
\end{multline}
is true for any smooth function $H=H(t,{q},{p})$. 
\end{lemma}

\begin{remark}
The identity~(\ref{ident}) can also be presented as
\begin{equation*}
 \nu\dot{q}  +  p D(\eta)  -  X(H)  -  HD(\xi)
\\
\equiv
\xi    {\delta    \tilde{H}   \over \delta   t}
+     \eta  {\delta  \tilde{H}   \over  \delta     q}
+\nu    {\delta   \tilde{H}   \over  \delta     p}
+ D \left[p\eta-\xi H\right].
\end{equation*}
\end{remark}

The identity~(\ref{ident})  was called the \textit{Hamiltonian
identity} as a version of the  Noether
identity~\cite{bk:DorodnitsynKozlov}. It leads to the following
result.

\begin{theorem} \label{firstintegral}
{\bf (Noether's theorem)}
The canonical Hamiltonian equations~(\ref{canonical}) possess a first integral of the form
\begin{equation}  \label{integral}
I=p\eta-\xi H
\end{equation}
if and only if the Hamiltonian function is invariant with respect
to the operator~(\ref{symmetry_H}) on the solutions of the equations~(\ref{canonical}).
\end{theorem}

Theorem~\ref{firstintegral}  corresponds to the strong version of the
Noether theorem (i.e., necessary and sufficient condition) for
invariant Lagrangians and Euler-Lagrange equations~\cite{bk:Ibragimov[1983]}. 
In the paper~\cite{bk:DorodnitsynKozlov}, the necessary and sufficient condition of the invariance of canonical Hamiltonian equations is also established.

\begin{theorem} \label{invarianceofvariations_b}
Canonical Hamiltonian equations~(\ref{canonical}) are invariant with respect to the symmetry~(\ref{symmetry_H})
{if and only if} the following conditions are true (on the solutions
of the canonical Hamiltonian equations)
\begin{subequations}  
\begin{gather}  
\label{new1}
\left.
\frac{\delta}{\delta p}\left({\nu}\dot{q}+pD({\eta})-X(H)-HD(\xi)\right)
\right|_{\dot{q} =H_{p},\ \dot{p}=-H_{q}}
=0,
\\
 \label{new2}
\left.
\frac{\delta}{\delta q}\left({\nu}\dot{q}+pD({\eta})-X(H)-HD(\xi)
\right)\right|_{\dot{q}=H_{p},\ \dot{p}=-H_{q}}
=0.
\end{gather}
\end{subequations}
\end{theorem}


\section{Variational framework to DODEs}
\label{section_delay_ODEs}

\subsection{Lagrangian approach}

In the recent paper~\cite{bk:DorodnitsynKozlovMeleshko[2023]},
there were considered  variations of  first-order functionals with one delay
\begin{subequations}      \label{functional_delay_L}
\begin{gather}       \label{functional_def}
{\cal L} = \int_{a}^{b}
{ L}(t, t^-,q,q^-,\dot{q},\dot{q}^-)  dt,
\\
 \label{delay_equation}
t - t^- = \tau,
\qquad
\tau = \mbox{const}.
\end{gather}
\end{subequations}
Here, we use the left shift  operator  $S_{-}  $ which is defined for a function
$ f = f ( t ) $ as
\begin{equation}   \label{shiftsB}
f^- = S_- (f)   = f(t - \tau  ).
\end{equation}
Analogously,   $S_{+}  $ is the right shift operator,
defined for a function $ f = f ( t ) $ as
\begin{equation}   \label{shifts}
f^+ = S_+ (f)   = f(t+\tau  ).
\end{equation}

Variations of the dependent variable $q$ and the independent variable  $t$
taken on the interval $ [ a, b- \tau] $
lead to the  equations
\begin{subequations}   \label{variational_L}
\begin{gather}
\label{variational_q}
{\frac{\delta  {L} }{\delta q}}
=
\frac{\partial { L}}{\partial q}
+ \frac{\partial { L^+}}{\partial q}
- {\DD}  \left( \frac{\partial { L}}{\partial \dot{q}}
 +   \frac{\partial { L^+}}{\partial \dot{q}} \right)
=0,
\\
  \label{variational_t}
{\frac{\delta  {L} }{\delta t}}
  = {  \partial { L} \over \partial t}
+ { \partial { L} ^+ \over \partial t }
+  {\DD} \left( \dot{q}  { \partial { L} \over \partial \dot{q} }
+ \dot{q}  { \partial { L} ^+ \over \partial \dot{q} }
-  { L} \right)
 = 0,
\end{gather}
 \end{subequations}
where
  $  L^+ = { L} (t^+,t, q^+,q,\dot{q}^+,\dot{q} ) $
is  the shifted to the right Lagrangian $ L = { L} (t,t^-, q,q^-,\dot{q},\dot{q}^ -)  $.
Here,  the total differentiation operator is acting on the variable in three points  $t$, $ t^-$ and $t^+$
\begin{equation*}    
D
=
\frac{\partial {}}{\partial t}
+\dot{q}\frac{\partial { }}{\partial q}
+ \ddot{q} \frac{\partial}{\partial\dot{q}}
+ \cdots
+ \frac{\partial {}}{\partial t^-}
+\dot{q}^-\frac{\partial { }}{\partial q^-}
+ \ddot{q} ^- \frac{\partial}{\partial\dot{q}^-}
+ \cdots
+ \frac{\partial {}}{\partial t^+}
+\dot{q}^+\frac{\partial { }}{\partial q^+}
+ \ddot{q}^+ \frac{\partial}{\partial\dot{q}^+}
+ \cdots
\end{equation*}

Equations~(\ref{variational_L}) are the vertical
(for variations of the dependent variable)
and the horizontal
(for variations of the independent variable)
variational equations,
respectively.
The variational  equation~(\ref{variational_q})  is known since Elsgolts~\cite{bk:Elsgolts[1955]}
(see also~\cite{Elsgolts2, Elsgolts3}).
We call it the {\it Elsgolts equation}.
This equation is a delay analog of the Euler-Lagrange equation~(\ref{Euler_Lagrange}).

For variations along group orbits corresponding to generators of the form~(\ref{symmetry_L}),
we obtain the variational equation
\begin{equation}         \label{quasi}
F_L
= \xi   { \delta {L}  \over \delta t }
+  \eta    { \delta {L}  \over \delta q }
= 0.
\end{equation}
which depends on the coefficients of the generator.
This equation is called the {\it local extremal equation}.
It contains variational equations~(\ref{variational_L})   as particular cases.
The local extremal equation is a second-order DODEs with two delays.
It is considered with the delay equation
\begin{equation}   \label{delay_eq}
t ^+   - t = t - t^- = \tau,
\qquad
\tau = \mbox{const}.
\end{equation}

The system~(\ref{quasi}),(\ref{delay_eq}) should be supplemented with initial conditions
on the interval $ [ a - \tau, a +  \tau] $.
Note that if we consider both variational equations~(\ref{variational_L})
with the delay equation~(\ref{delay_eq}),
we get an {over}determined system.
In the present paper, we are interested in developing Hamiltonian analogs of DODEs
~(\ref{variational_q}),~(\ref{variational_t}) and~(\ref{quasi}).

Invariance of a system which consists of a second-order DODE
with two delays (such as  equation~(\ref{variational_q}),~(\ref{variational_t})  or~(\ref{quasi}))
and the delay equation~(\ref{delay_eq})
was examined in~\cite{bk:DorodnitsynKozlovMeleshko[2023]}.
It was established  that the invariance of the delay equation
restricts the symmetry coefficient $  \xi (t,q) $ by the condition
\begin{equation}     \label{xi_condition}
D ( \xi ) = D ( \xi ^- ), 
\end{equation}
which should hold on the solutions of the considered system. 
It follows that  $  \xi (t,q) =  \xi (t) $.
The general solution of equation~(\ref{xi_condition}) is
\begin{equation}     \label{xi_we_can}
 \xi (t) = \alpha t  +  f(t),
\end{equation}
where $ \alpha $ is a constant and $f(t) $  is a periodic function with period $\tau $.
For symmetry applications~\cite{bk:DorodnitsynKozlovMeleshko[2023]},
we proceed with
\begin{equation}    \label{xi_we_take}
\xi (t) = A t + B,
\end{equation}
where $A$  and $B$ are arbitrary constants.

\subsection{Hamiltonian approach}

Similarly to the continuous case, one can consider a delay analog
of the functional~(\ref{functional_H}).  
We introduce the
Hamiltonian functional with one delay 
\begin{equation}       \label{functional_H_delay}
 {\cal H}
=
\int  _a ^b
p^{-}(\alpha_{1} d \dot{q}+\alpha_{2} d \dot{q}^{-})
+p(\alpha_{3} d \dot{q}+\alpha_{4} d \dot{q}^{-})
-H(t,t^{-},q,q^{-},p,p^{-}) \ d t,
\end{equation}
where   $ \alpha_{i} $, $ i = 1,2,3,4$ are some constants. One can
consider more complicated cases of the coefficients $ \alpha_{i} $.
{Non}constant   $ \alpha_{i} $  are treated in
Appendix~\ref{section_algorithm}.

For the variations of the dependent variables $p$  and $q$
and the independent variable $t$ taken on the interval $    [ a. b - \tau] $,
we obtain equations  
\begin{subequations}        \label{delay_variation}
\begin{gather}
 \label{delay_variation_p}
{ \delta  \tilde{H}  \over \delta p }
=
\alpha_{1}\dot{q}^{+}
+(\alpha_{2}+\alpha_{3})\dot{q}
+\alpha_{4}\dot{q}^{-}
-\frac{\partial }{\partial{p}}   (H+H^{+})
= 0,
\\
 \label{delay_variation_q}
{ \delta  \tilde{H}  \over \delta q }
= - \left(
\alpha_{4}\dot{p}^{+}
+(\alpha_{2}+\alpha_{3})\dot{p}
+\alpha_{1}\dot{p}^{-}
+\frac{\partial  }{\partial q} (H+H^{+})
\right)
=  0,
\\
  \label{delay_variation_t}
{ \delta  \tilde{H}  \over \delta t }
=
    D  [\alpha_{2}(p\dot{q}-p^{-}\dot{q}^{-})+\alpha_{4}(p^{+}\dot{q}-p\dot{q}^{-}) ]
    +D(H)
   - \frac{\partial  }{\partial t}   (H+H^{+})
= 0,
\end{gather}
\end{subequations}
where
\begin{equation}       \label{H_tilde}
\tilde{H}
=
p^{-}(\alpha_{1}\dot{q}+\alpha_{2}\dot{q}^{-})
+p(\alpha_{3}\dot{q}+\alpha_{4}\dot{q}^{-})
-H(t,t^{-},q,q^{-},p,p^{-})
\end{equation}       
is the function corresponding to the integrand of
functional~(\ref{functional_H_delay}), and  the variational
operators are
\begin{subequations}   \label{varoperator1_delay}
\begin{gather}
\label{varoperator1_delay_p}
\frac{\delta}{\delta p}
=
\frac{\partial}{\partial p}
- D\frac{\partial}{\partial\dot{p}}
+ S_+ \left(
\frac{\partial}{\partial p^- }
- D\frac{\partial}{\partial\dot{p}^- }
\right),
\\
\label{varoperator1_delay_q}
\frac{\delta}{\delta q}
=
\frac{\partial}{\partial q}
- D \frac{\partial}{\partial\dot{q}}
+ S_+ \left(
\frac{\partial}{\partial q ^- }
- D \frac{\partial}{\partial \dot{q} ^- }
\right),
\\
\label{varoperator1_delay_t}
\frac{\delta}{\delta t}
=
\frac{\partial}{\partial t}
+  D  \left(     \dot{q}   \frac{\partial}{\partial\dot{q}  }
  +  \dot{p}   \frac{\partial}{\partial\dot{p}  }      \right)
+ S_+ \left(
\frac{\partial}{\partial t^-}
+  D  \left(     \dot{q}  ^-  \frac{\partial}{\partial\dot{q}  ^- }
  +  \dot{p}  ^-   \frac{\partial}{\partial\dot{p} ^-  }      \right)
\right)
- D.
\end{gather}
\end{subequations}
Here, the operator of total differentiation   $D$  is
\begin{equation*}  
 D
=\frac{\partial}{\partial t}
+\dot{q}\frac{\partial}{\partial q}
+\dot{p}\frac{\partial}{\partial p}
+  \cdots
+ \frac{\partial}{\partial t^-}
+\dot{q}^- \frac{\partial}{\partial q^-}
+\dot{p}^- \frac{\partial}{\partial p^-}
+  \cdots
+ \frac{\partial}{\partial t^+}
+\dot{q}^+ \frac{\partial}{\partial q^+}
+\dot{p}^+ \frac{\partial}{\partial p^+}
+  \cdots
 \end{equation*}
The first and second operators perform the "vertical variations"
(i.e., the variations with respect to the dependent variables $p$ and
$q$), the last one takes the "horizontal variation" (the variation
with respect to the independent variable $t$).

\begin{remark}
In contrast to the Hamiltonian ODEs
(see Remark~\ref{the_third_equation}),
the equation~(\ref{delay_variation_t})
does not hold on the solutions of the equations~(\ref{delay_variation_p})  and~(\ref{delay_variation_q}).
\end{remark}

For the variations along group orbits corresponding to the
generator~(\ref{symmetry_H}), we obtain the variational equation
\begin{equation}       \label{quasi_H}
\xi
{ \delta  \tilde{H}  \over \delta t }
+
\eta
{ \delta  \tilde{H}  \over \delta q }
+ \nu
{ \delta  \tilde{H}  \over \delta p }
= 0.
\end{equation}
This local extremal equation is the Hamiltonian analog of~(\ref{quasi}).

We  consider the system of equations
\begin{subequations}         \label{delay_canonical}
\begin{gather}    \label{delay_canonical_p}
\alpha_{1}\dot{q}^{+}
+(\alpha_{2}+\alpha_{3})\dot{q}
+\alpha_{4}\dot{q}^{-}
=
\frac{\partial }{\partial{p}}   (H+H^{+})
\\
     \label{delay_canonical_q}
\alpha_{4}\dot{p}^{+}
+(\alpha_{2}+\alpha_{3})\dot{p}
+\alpha_{1}\dot{p}^{-}
=
-  \frac{\partial }{\partial q}  (H+H^{+}),
\end{gather}
\end{subequations}
given by~(\ref{delay_variation_p}) and~(\ref{delay_variation_q})
as the {\it delay  canonical Hamiltonian equations},
i.e., the delay analog of the canonical Hamiltonian equations~(\ref{canonical}).  
Shortly, it can be presented as
\begin{equation}    \label{delay_canonical_short}
{ \delta  \tilde{H} \over \delta p }  =0,
\qquad
{ \delta  \tilde{H}  \over \delta q }  = 0.
\end{equation}
These equations are first-order DODEs with two delays. 
They are supplemented by the delay equation~(\ref{delay_eq}). 
Note that all three variational equations  (\ref{varoperator1_delay}) 
supplemented by the delay equation~(\ref{delay_eq}) form an {over}determined system.

\section{Compatibility of Hamiltonian approach with Lagrangian approach}
\label{section_Hamiltonian_formalism}

 In the classical Hamiltonian theory (see, for example,~\cite{Gelfand}), 
the variables $q$ and $p$ are named canonical, 
if the system of equations for them is equivalent to the Euler-Lagrange variational equation. 
We are following the same strategy for  the delay Hamiltonian equations 
and considering the way from the Lagrangian approach to the Hamiltonian approach 
by means of a delay version of the Legendre transformation.


\subsection{Delay Legendre transformation}

We consider the delay analog of the Legendre transformation 
\begin{equation}    \label{delay_Legendre}
H(t,t^{-},q,q^{-},p,p^{-})
=
p^{-}(\alpha_{1}\dot{q}+\alpha_{2}\dot{q}^{-})
+p(\alpha_{3}\dot{q}+\alpha_{4}\dot{q}^{-})
-L  (t,t^{-},q,q^{-},\dot{q},\dot{q}^{-}),
\end{equation}
where $\dot{q} $  and  $\dot{q} ^- $ in the right-hand side are to be substituted as found from
\begin{equation}    \label{234gf}
 \alpha_{1}  {p^- }   +  \alpha_{3}  {p}
 =\frac{\partial L}{\partial\dot{q}},
\qquad
  \alpha_{2}  {p^- }   +  \alpha_{4}  {p}
 =\frac{\partial L}{\partial\dot{q}^{-}}.
\end{equation}
We also obtain relations
\begin{equation}
 \label{conditions_2}
\alpha_{3}\dot{q}+\alpha_{4}\dot{q}^{-}
= \frac{\partial H}{\partial p },
\qquad
\alpha_{1}\dot{q}+\alpha_{2}\dot{q}^{-}
= \frac{\partial H}{\partial p ^- },
\end{equation}
and
\begin{equation}
 \label{conditions_3}
\frac{\partial H}{\partial q }   =  -  \frac{\partial L}{\partial q },
\qquad
\frac{\partial H}{\partial q ^- }   =  -  \frac{\partial L}{\partial q^-  },
\qquad
\frac{\partial H}{\partial t }   =  -  \frac{\partial L}{\partial t },
\qquad
\frac{\partial H}{\partial t^-  }   =  -  \frac{\partial L}{\partial t^- }.
\end{equation}

\subsection{Are any constants $\alpha_i$ good for the delay Legendre transformation?}
\label{Preliminary}

In the following example, we choose some constants $\alpha_i$ and examine the Legendre transformation.

Let the following Legendre relation link the Lagrange and the
Hamilton functions 
\begin{equation}  \label{123}
H(t,t^{-},q,q^{-},p,p^{-})
= p\dot{q}+p^{-}\dot{q}^{-}
-L  (t,t^{-},q,q^{-},\dot{q},\dot{q}^{-}),
\end{equation}
where       $ \dot{q}$ and  $ \dot{q}^{-}$  can be found from
\begin{equation}   \label{234}
{p}=\frac{\partial L}{\partial \dot{q}},
\qquad
{p^{-}}=\frac{\partial L}{\partial  \dot{q}^{-}}
\end{equation}
and then substituted into the right-hand side of the relation to
find the Hamiltonian function from the Lagrangian.

Supposing that $p=S_{+} (p^- )$,we immediately get 
\begin{equation*}  
{p}
= S_{+} \left(\frac{\partial L}{\partial\dot{q}^- }\right)
= \frac{\partial L^{+}}{\partial\dot{q}}.
\end{equation*}
It follows that the Lagrangian should satisfy
\begin{equation*}
\frac{\partial L^{+}}{\partial\dot{q}}=\frac{\partial L}{\partial\dot{q}}.
\end{equation*}
Evidently,  any Lagrangian does not obey this property. For example
the Lagrangian
\begin{equation}  \label{353}
L
=\dot{q}\dot{q}^{-}+({\dot{q}}^{-})^{2}-qq^{-}
\end{equation}
does not satisfy it.

Thus, we see that the Legendre transformation should be adapted to
the form of a delay Lagrangian.

We are interested in delay canonical Hamiltonian
equations~(\ref{delay_canonical}), which correspond to Elsgolts equations~(\ref{variational_q}). 
Starting with a delay Lagrangian $L$, we obtain the corresponding delay Hamiltonian $H$. 
The delay Legendre transformation~(\ref{delay_Legendre}) gives the
correspondence between the Lagrangian and the Hamiltonian. 
We impose a {\it compatibility condition}\footnote{ Imposing this condition, 
we avoid contradiction with the Legendre relations in a given point  with that in a shifted point,
and avoid introducing a double set of variables $p$ as it was done in~\cite{bk:Zhang}.}, 
which requires that the variables $p$ and $p^- $ expressed from the  Legendre transformation
via the variables $\dot{q}$ and  $\dot{q}^-$ satisfy
\begin{equation}    \label{requirement_p}
p=S_{+} (p^- ). 
\end{equation}

\subsection{Delay Legendre transformation with compatibility condition}



We consider Lagrangians  to be quadratic in the derivatives.
Namely, the Lagrangians  of the form
\begin{equation}     \label{quadratic_L}
{L}
=  { \alpha  \over 2 }    {\dot{q}}^{2}
+  \beta{\dot{q}\dot{q}}^{-}
+  { \gamma    \over 2 } (\dot{q}^{-})^{2}
-   \phi(q,q^{-}),
\end{equation}
where $\alpha$, $\beta \neq 0$ and $\gamma$ are some constants 
(a more general form of Lagrangians is examined in Appendix~\ref{section_algorithm}). 
The Elsgolts variational equation for  this Lagrangian takes the form
\begin{equation}     \label{Elsgolts_variational_equation}
{\delta {L} \over \delta q }
=     -  \left(
\beta{\ddot{q}}^{+}
 +  (\alpha+\gamma)\ddot{q}
 +   \beta{\ddot{q}}^{-}
 +   \frac{\partial}{\partial q}     (\phi+\phi^{+})
\right)
=0.
\end{equation}

\subsubsection{{Non}degenerate case $ \alpha \gamma -{\beta}^{2} \neq 0 $}

Applying the equations~(\ref{234gf}) to the quadratic Lagrangian~(\ref{quadratic_L}),
we obtain
\begin{equation}   \label{relations}
\alpha_{1}p^{-}+\alpha_{3}p
=
\alpha\dot{q}+\beta\dot{q}^{-},
\qquad
\alpha_{2}p^{-}+\alpha_{4}p
=
\beta\dot{q}+\gamma\dot{q}^{-}.
\end{equation}
We require the conditions
\begin{subequations}
\begin{gather}
 \label{coefficient_condition_1}
\alpha_{1} \alpha_{4} -   \alpha_{2} \alpha_{3} \neq 0,
\\
 \label{coefficient_condition_2}
\alpha \gamma -{\beta}^{2} \neq 0
\end{gather}
\end{subequations}
so that the  relations~(\ref{relations}) can be resolved   for $ p$  and $p^- $  as well as for $ \dot{q}$ and  $ \dot{q} ^- $.
Lagrangians satisfying these conditions are called {non}degenerate.
We obtain
\begin{subequations}   \label{eq:mar10.10}
\begin{gather}
{\displaystyle
p
=
\frac{  \beta \alpha_{1}  -   \alpha\alpha_{2}}{ \alpha_{1}\alpha_{4} - \alpha_{2}\alpha_{3}}  \dot{q}
+
\frac{ \gamma \alpha_{1}  - \beta\alpha_{2}}{ \alpha_{1}\alpha_{4} - \alpha_{2}\alpha_{3}  }{\dot{q}}^{-} },
\\
{\displaystyle
p^{-}
=
\frac{  \alpha\alpha_{4}-\beta\alpha_{3} }{\alpha_{1}\alpha_{4}-\alpha_{2}\alpha_{3}} \dot{q}
+ \frac{  \beta\alpha_{4}- \gamma\alpha_{3} }{\alpha_{1}\alpha_{4}-\alpha_{2}\alpha_{3}}  {\dot{q}}^{-} }
\end{gather}
\end{subequations}
and
\begin{subequations}   \label{eq:mar10/9}
\begin{gather}
{\displaystyle
\dot{q}
=
\frac{\gamma\alpha_{3}-\beta{\alpha}_{4}}{ \alpha \gamma-{\beta}^{2}}p
+
\frac{\gamma\alpha_{1}-\beta{\alpha}_{2}}{ \alpha \gamma-{\beta}^{2}}p^{-} },
\\
{\displaystyle
\dot{q}^{-}
=
\frac{\alpha\alpha_{4}-\beta{\alpha}_{3}}{ \alpha \gamma-{\beta}^{2}}p
+
\frac{\alpha\alpha_{2}-\beta{\alpha}_{1}}{ \alpha \gamma-{\beta}^{2}}p^{-}  }.
\end{gather}
\end{subequations}



The compatibility condition~(\ref{requirement_p}) leads to the relation
\begin{equation*}
\frac{  \beta \alpha_{1}  -   \alpha\alpha_{2}}{ \alpha_{1}\alpha_{4} - \alpha_{2}\alpha_{3}}  \dot{q}
+
\frac{ \gamma \alpha_{1}  - \beta\alpha_{2}}{ \alpha_{1}\alpha_{4} - \alpha_{2}\alpha_{3}  }{\dot{q}}^{-}
=
\frac{  \alpha\alpha_{4}-\beta\alpha_{3} }{\alpha_{1}\alpha_{4}-\alpha_{2}\alpha_{3}} \dot{q}^+
+ \frac{  \beta\alpha_{4}- \gamma\alpha_{3} }{\alpha_{1}\alpha_{4}-\alpha_{2}\alpha_{3}}  {\dot{q}}.
\end{equation*}
Splitting this equation for $  \dot{q}^+ $, $  \dot{q}$,  and $  \dot{q}^. $,
one derives
\begin{equation}    \label{coefficients}
\alpha_{2}=\frac{\gamma}{\beta}  \alpha_{1},
\qquad
\alpha_{3}=\frac{\alpha}{\beta}   \alpha_{1},
\qquad
\alpha_{4}=\alpha_{1} \neq 0.
\end{equation}
We remark that for these coefficients, condition~(\ref{coefficient_condition_1}) follows from condition~(\ref{coefficient_condition_2}).
Thus, we can define the {non}degenerate Lagrangians as Lagrangians satisfying~(\ref{coefficient_condition_2}).

For the coefficients~(\ref{coefficients}), we obtain
\begin{equation}     \label{relations_p}
p =   {\beta \over \alpha_1 }    \dot{q},
\qquad
p^- =    {\beta \over \alpha_1 }    \dot{q}^-.
\end{equation}
The Hamiltonian given by~(\ref{delay_Legendre}) takes the form
\begin{equation}  \label{Hamiltonian}
H=\frac{\alpha_{1}^{2}}{\beta^{2}}
\left(  { \alpha  \over 2 } p{}^{2}  +\beta pp^{-}+ { \gamma \over 2 }  (p^{-}) ^{2} \right)
+ \phi ( q, q^- ).
\end{equation}
For this Hamiltonian,
the  delay Hamiltonian equations~(\ref{delay_canonical}) with the  coefficients~(\ref{coefficients})
can be presented as
\begin{subequations}     \label{eq:GenHamEq_d}
\begin{gather}
{\displaystyle
\alpha_{1}\left(\dot{q}^{+}+ \frac{\alpha + \gamma }{\beta}\dot{q}+\dot{q}^{-}\right)
= \frac{\alpha_{1}^{2}}{\beta}
\left(   p^{+}  +    { \alpha    + \gamma  \over \beta} p  +   p^{-}   \right)  },
\\
{\displaystyle  \alpha_{1}\left(\dot{p}^{+}+ \frac{\alpha+ \gamma}{\beta}\dot{p}+\dot{p}^{-}\right)
=-\frac{\partial}{\partial q} }  ( \phi  + \phi ^{+} ).
\end{gather}
\end{subequations}
They provide a decomposition of  the second-order Elsgolts equation~(\ref{Elsgolts_variational_equation})
into two first-order DODEs with two delays.
A convenient choice  $ \alpha _1 = \beta $ simplifies the relations~(\ref{relations_p}),
the Hamiltonian~(\ref{Hamiltonian}),
and the delay canonical Hamiltonian equations~(\ref{eq:GenHamEq_d}).

\begin{example}  \label{Non_degenerate case}
{\bf Delay  ({non}degenerate) harmonic oscillator}

We consider the Lagrangian
\begin{equation}
{L} (t,t^{-},q,q^{-},\dot{q},\dot{q}^{-})=\dot{q}\dot{q}^{-}-qq^{-}.
\end{equation}
It  yields the Elsgolts equation
\begin{equation}    \label{Elsgolts_eq_example_1}
\frac{\delta {L} }{\delta q}
=-(\ddot{q}^{+}+\ddot{q}^{-}+q^{+}+q^{-})=0.
\end{equation}
The Lagrangian is {non}degenerate
since the coefficients $\alpha =0$, $ \beta =1 $,  and $\gamma = 0$ satisfy~(\ref{coefficient_condition_2}).

Following~(\ref{coefficients}), we take
\begin{equation*}
\alpha_1 = 1,
\qquad
\alpha_2 = 0,
\qquad
\alpha_3 = 0,
\qquad
\alpha_4 = 1.
\end{equation*}
For these coefficients,
the relations~(\ref{relations_p}) take the form
\begin{equation*}
p = \dot{q},
\qquad
 p ^- = \dot{q} ^-.
\end{equation*}

We obtain the Hamiltonian function
\begin{equation}
H
= p^{-}\dot{q}+p\dot{q}^{-}-{L}
= p^{-}\dot{q}+p\dot{q}^{-}-\dot{q}\dot{q}^{-}+qq^{-}
=pp^{-}+qq^{-}
\end{equation}
and the delay canonical Hamiltonian equations
\begin{subequations}
\begin{gather}
\dot{q}^{+}+\dot{q}^{-}
=p^{+}+p^{-},
\\
\dot{p}^{+}+\dot{p}^{-}
=- q^{+} - q^{-}.
\end{gather}
\end{subequations}
One can easily verify that these Hamiltonian equations are equivalent
to the Elsgolts equation~(\ref{Elsgolts_eq_example_1}).
\end{example}

\subsubsection{Degenerate case  $ \alpha \gamma -{\beta}^{2} =  0 $}

The degenerate Lagrangians~(\ref{quadratic_L})
with coefficients   $\alpha $, $ \beta$, and $\gamma$  satisfying
\begin{equation}   \label{degenerate_coefficients}
\alpha \gamma  -   {\beta}^{2} =  0
\end{equation}
should be considered separately.

For convenience, we restrict ourselves to the coefficients
\begin{equation*}
\alpha  > 0,
\qquad
\gamma  > 0,
\qquad
 \beta  =  \sqrt{\alpha  \gamma} > 0.
\end{equation*}
The other cases of~(\ref{degenerate_coefficients}) are examined similarly.
The  Lagrangian function can be rewritten as
\begin{equation}   \label{L_rewritten}
L
=  { \alpha  \over 2 }    {\dot{q}}^{2}
+     \sqrt{\alpha  \gamma} {\dot{q}\dot{q}}^{-}
+  { \gamma    \over 2 } (\dot{q}^{-})^{2}
-   \phi(q,q^{-})
=
  { 1 \over 2 }  (   \sqrt{\alpha}  \dot{q} +   \sqrt{\gamma} \dot{q} ^{-} )^{2}
- \phi(q, q^{-} ).
\end{equation}
The Elsgolts equation  is
\begin{equation}          \label{Elsgolts_rewritten}
{\delta  {L} \over \delta q }
=     -  \left(
\sqrt{\alpha \gamma }  {\ddot{q}}^{+}
 +  (\alpha+\gamma)\ddot{q}
 +   \sqrt{\alpha  \gamma } {\ddot{q}}^{-}
 +   \frac{\partial}{\partial q}     (\phi+\phi^{+})
\right)
=0.
\end{equation}

Equations~(\ref{234gf}) become
\begin{equation}  \label{eq:apr3.1}
\alpha_{1}p^{-}+\alpha_{3}p
= \alpha  \dot{q}   + \sqrt{\alpha  \gamma}    \dot{q} ^{-},
\qquad
\alpha_{2}p^{-}+\alpha_{4}p
=    \sqrt{\alpha  \gamma}   \dot{q} +  \gamma  \dot{q} ^{-}.
\end{equation}
Due to~(\ref{degenerate_coefficients}),
the right-hand sides of these equations are proportional.
From these  equations,  one obtains
\begin{equation*}
{ 1 \over \sqrt{\alpha} }
(  \alpha_{1}p^{-}+\alpha_{3}p )
=   {  1  \over \sqrt{\gamma} }  ( \alpha_{2}p^{-}+\alpha_{4}p  ).
\end{equation*}
We assume
\begin{equation}  \label{degenerate_coefficients_1}
 \alpha_{2} = \sqrt{ \gamma \over \alpha }  \alpha_{1},
\qquad
\alpha_{3} = \sqrt{ \alpha  \over \gamma }  \alpha_{4}.
\end{equation}
Eliminating   $ \alpha_{2} $ and $ \alpha_{3} $ from both equations~(\ref{eq:apr3.1}),
we obtain two proportional equations. 
Thus, the system  (\ref{eq:apr3.1}) is equivalent to one equation 
\begin{equation}  \label{eq_merged}
{ \alpha_{1}  \over   \sqrt{\alpha}  } p^{-}
+ {\alpha_{4}   \over   \sqrt{\gamma}  } p
= \sqrt{\alpha}     \dot{q}
+ \sqrt{\gamma}    \dot{q} ^{-}.
\end{equation}


Now, we can derive the delay Hamiltonian.
First, we use coefficients~(\ref{degenerate_coefficients_1})
in the delay Legendre transformation~(\ref{delay_Legendre})
\begin{multline*}
H
=p^{-} \left(\alpha_{1}  \dot{q} + \sqrt{ \gamma \over \alpha }  \alpha_{1}  \dot{q} ^{-} \right)
+p  \left(  \sqrt{ \alpha  \over \gamma }  \alpha_{4}  \dot{q}+\alpha_{4}  \dot{q}^{-} \right)
-  L
\\
=p^{-}    { \alpha_{1}  \over   \sqrt{\alpha}  }
(   \sqrt{\alpha}  \dot{q} + \sqrt{\gamma}    \dot{q} ^{-} )
+p    {\alpha_{4}   \over   \sqrt{\gamma}  }
(  \sqrt{\alpha}   \dot{q}+   \sqrt{\gamma}    \dot{q}^{-} )
-  L
\\
=  \left(
  { \alpha_{1}  \over   \sqrt{\alpha}  }    p^-
+    {\alpha_{4}   \over   \sqrt{\gamma}  }  p
\right)
(   \sqrt{\alpha}  \dot{q} + \sqrt{\gamma}    \dot{q} ^{-} )
-  L.
\end{multline*}
Then, using relation~(\ref{eq_merged})
and  the Lagrangian,
rewritten as given in~(\ref{L_rewritten}),
we obtain the Hamiltonian
\begin{equation}
H =\frac{1}{2}   \left(
{ \alpha_{4}  \over \sqrt{\gamma}  } p
+ {  \alpha_{1 }       \over \sqrt{\alpha}  }  p ^-
\right)^{2}
+  \phi ( q, q^-).
\end{equation}

The Hamiltonian provides the delay canonical Hamiltonian equations
\begin{subequations}
\begin{gather}
\label{eq:Ham_p}
\alpha_{1} \dot{q}^{+}
+  \left(   \sqrt{\gamma \over \alpha}  \alpha_{1}
+   \sqrt{\alpha \over \gamma }  \alpha_{4}  \right)  \dot{q}
+\alpha_{4} \dot{q}^{-}
=
 {\alpha_4  \over \sqrt{\gamma} }
\left(    {\alpha_1  \over \sqrt{\alpha} }    p^-
+ {\alpha_4  \over \sqrt{\gamma} }   p \right)
  +
{\alpha_1  \over \sqrt{\alpha} }
\left(    {\alpha_1  \over \sqrt{\alpha} }    p
+ {\alpha_4  \over \sqrt{\gamma} }   p^+  \right),
\\
 \label{eq:Ham_q}
\alpha_{4} \dot{p}^{+}
+ \left(  \sqrt{\gamma \over \alpha}   \alpha_{1}
+   \sqrt{\alpha \over \gamma }  \alpha_{4}  \right)   \dot{p}
+\alpha_{1}\dot{p}^{-}
=
-  \frac{\partial }{\partial q}  (\phi+\phi^{+}).
\end{gather}
\end{subequations}
We require these Hamiltonian equations to be equivalent to the Elsgolts equation~(\ref{Elsgolts_rewritten}). 
Assuming  $ \alpha _1  \neq 0$ and $ \alpha _4  \neq 0$, 
we express $p ^{+}$ from equation (\ref{eq:Ham_p}), differentiate, 
and substitute  $\dot{p}^{+}$ into  equation (\ref{eq:Ham_q}). 
We obtain 
\begin{equation*}
\sqrt{\alpha \gamma}  \ddot{q}^{+}
+ \left(     \gamma  + \alpha  {\alpha _4   \over \alpha _1 } \right)   \ddot{q}
+  \sqrt{\alpha \gamma}      {\alpha _4   \over \alpha _1 }  \ddot{q}^{-}
+  (    \alpha _1 -   \alpha _4  )
 \left(  
 \sqrt{\alpha \over  \gamma}   {\alpha _4   \over \alpha _1 }     
 \dot{p}  
 +   \dot{p}   ^-    
\right)
=
-  \frac{\partial }{\partial q}  (\phi+\phi^{+}). 
\end{equation*}
To make this equation equivalent to the  Elsgolts equation, 
we can set 
\begin{equation*}
\alpha _4 =  \alpha _1   . 
\end{equation*}
Thus we confirmed that coefficients~(\ref{coefficients}) are also suitable
for the case of the degenerate Lagrangians.

We conclude that in both {non}degenerate and degenerate cases of quadratic Lagrangians  (\ref{quadratic_L}), 
we obtain coefficients $ \alpha _i$ which are also given by~(\ref{coefficients}).
It is worth mentioning that one can derive the coefficients~(\ref{coefficients})
using and an alternative approach presented in Appendix~\ref{Alternative_approach}.

\begin{example}
\label{Degenerate case}
{\bf Delay (degenerate) oscillator}

Consider the Lagrangian
\begin{equation}
{L} (t,t^{-},q,q^{-},\dot{q},\dot{q}^{-})
= {  1 \over 2 } (\dot{q}+\dot{q}^{-})^{2}
-  \phi(q, q^{-}).
\end{equation}
The Elgolts equation is
\begin{equation}    \label{eq:apr3_Els}
\frac{\delta {L}}{\delta q}
=- \left(
\ddot{q}^{+}+2\ddot{q}+\ddot{q}^{-}
+   {\partial \over \partial q} (\phi+\phi^{+})
\right)
=0.
\end{equation}
Because the coefficients $\alpha =1$, $ \beta =1 $,  and $\gamma = 1$ do not satisfy the condition~(\ref{coefficient_condition_2}),
we have a degenerate case.

Using~(\ref{coefficients}), we choose
\begin{equation*}
\alpha_1 = 1,
\qquad
\alpha_2 = 1,
\qquad
\alpha_3 = 1,
\qquad
\alpha_4 = 1.
\end{equation*}
The system~(\ref{relations}) is degenerate.
It is equivalent to one equation
\begin{equation*}
p + p^- = \dot{q}  +  \dot{q} ^-.
\end{equation*}
However, this relation is sufficient to derive the delay Hamiltonian.

We get the Hamiltonian function
\begin{multline}
H
=
p^{-}( \dot{q}+ \dot{q}^{-} )
+ p( \dot{q}+\dot{q}^{-} )
- L
\\
=
( p +  p^{-}  )  ( \dot{q}+ \dot{q}^{-} )
- {  1 \over 2 }  (\dot{q}+\dot{q}^{-})^{2}
+  \phi(q, q^{-})
=
\frac{1}{2}(  p + p^{-} )^{2}
+  \phi(q, q^{-}).
\end{multline}
It leads to the delay canonical Hamiltonian equations
\begin{subequations}
\begin{gather}
\dot{q}^{+}+2\dot{q}+\dot{q}^{-}
=
{p}^{+}+2{p}+{p}^{-},
\\
\dot{p}^{+}+2\dot{p}+\dot{p}^{-}
=
-  \frac{\partial }{\partial q}    (\phi+\phi^{+}),
\end{gather}
\end{subequations}
which are equivalent  to the Elsgolts equation~(\ref{eq:apr3_Els}).
\end{example}

\subsection{Reverse transformation: from Hamiltonian to Lagrangian}

The approach based on a compatibility  condition can also be used to obtain  Lagrangians
starting with Hamiltonians.
Given  a quadratic Hamiltonian function
\begin{equation}  \label{from_Hamiltonian}
H=
  { A  \over 2 } p{}^{2}  + B pp^{-}+ { C \over 2 }  (p^{-}) ^{2}
+ \phi ( q, q^- ),
\end{equation}
where  $A$, $B\neq 0 $ and $C$ are constant coefficients,
we employ the conditions~(\ref{conditions_2})
\begin{equation}   \label{relations_2}
\alpha_{3}\dot{q}+\alpha_{4}\dot{q}^{-}
= A p + B p ^-,
\qquad
\alpha_{1}\dot{q}+\alpha_{2}\dot{q}^{-}
= B p  + C p^-.
\end{equation}
From these equations,
we proceed similarly to the case of deriving a delay Hamiltonian from a delay Lagrangian.
We express $\dot{q}$ and $ \dot{q}  ^- $ with the help  $p$ and $p^-$
and impose   the compatibility condition
\begin{equation}    \label{requirement_q}
\dot{q}  =  S_+ (  \dot{q}  ^-).
\end{equation}
The analysis splits into two cases:
a) {non}degenerate case with $ A C - B^2 \neq 0 $
and
b) degenerate case with $ A C - B^2 = 0 $.
In both cases, we obtain the coefficients
\begin{equation}    \label{coefficients_2}
\alpha_{2}=\frac{C}{B}  \alpha_{1},
\qquad
\alpha_{3}=\frac{A}{B}   \alpha_{1},
\qquad
\alpha_{4}=\alpha_{1} \neq 0.
\end{equation}

For these coefficients, we get
\begin{equation}     \label{relations_qp}
 \dot{q} =  {B \over \alpha_1 }   p,
\qquad
 \dot{q} ^- =  {B \over \alpha_1 }   p  ^-.
\end{equation}
The corresponding Lagrangian function is
\begin{equation}     \label{to_Lagrangian}
L
=\frac{\alpha_{1}^{2}}{B ^{2}}
\left(  { A  \over 2 } \dot{q}  ^{2}  +  B  \dot{q} \dot{q} ^{-}+ { C  \over 2 }  (\dot{q} ^{-}) ^{2} \right)
-  \phi ( q, q^- ).
\end{equation}

The Hamiltonian~(\ref{from_Hamiltonian}) gives  the delay canonical Hamiltonian equations
\begin{subequations}        \label{delay_canonical_H}
\begin{gather}    \label{delay_canonical_p_H}
{ \alpha_{1} \over B}
\left(
B \dot{q}^{+}
+(  A + C   ) \dot{q}
+ B \dot{q}^{-}
\right)
=
 B p ^+ + ( A + C ) p + B p^-
\\
     \label{delay_canonical_q_H}
{ \alpha_{1} \over B}
\left(
B  \dot{p}^{+}
+(  A + C   )  \dot{p}
+ B\dot{p}^{-}
\right)
=
-  \frac{\partial }{\partial q}  (\phi+\phi^{+}).
\end{gather}
\end{subequations}
The Lagrangian~(\ref{to_Lagrangian}) provides  the Elsgolts equation
\begin{equation}        \label{delay_L_H}
{\delta {L} \over \delta q }
=     -  \left(
\frac{\alpha_{1}^{2}}{B ^{2}}
\left(
B {\ddot{q}}^{+}
 +  (A +  C ) \ddot{q}
 +  B {\ddot{q}}^{-}
\right)
 +   \frac{\partial}{\partial q}     (\phi+\phi^{+})
\right)
=0.
\end{equation}
Comparison shows that the delay canonical Hamiltonian equations~(\ref{delay_canonical_H})
and the Elsgolts equation~(\ref{delay_L_H}) are equivalent.
Note that a convenient choice  $ \alpha _1 = B $ makes simplification.

There is an example of $H$ considered without any $L$ in
Appendix~\ref{H_without_L}.


\section{Lie point symmetries and invariance}
\label{section_symmetries}

Let  group transformations  in the space $(t,{q},{p})$
be  provided by generators of the form~(\ref{symmetry_H})
To apply the generator~(\ref{symmetry_H})  to delay Hamiltonian equations,
we prolong it to all variables involved, namely
derivatives of $\dot{q} $ and $ \dot{p}$
and variables at the shifted points
$(t^{-},q^{-},  p^-, \dot{q}^-, \dot{p}^-) $
and
$(t^{+},q^{+}, p^+, \dot q^{+}, \dot{p}^+) $.
We obtain the prolonged operator
\begin{multline} \label{operator2}
{X}
={\xi}{\partial \over \partial t}
+ {\eta} {\partial  \over \partial q}
 + \nu  \frac{\partial}{\partial p}
+ {\zeta} _{\eta}   {\partial  \over \partial \dot{q}}
+ {\zeta} _{\nu}   {\partial  \over \partial \dot{p}}
\\
+ {\xi}^- {\partial \over \partial t^- }
+ {\eta}^-  {\partial  \over \partial q^- }
+ \nu ^-  \frac{\partial}{\partial p^-}
+ {\zeta} _{\eta}  ^-   {\partial  \over \partial \dot{q}^- }
+ {\zeta} _{\nu}  ^- {\partial  \over \partial \dot{p}^- }
\\
+ {\xi}^+ {\partial \over \partial t^+ }
+ {\eta}^+  {\partial  \over \partial q^+ }
+ \nu  ^+ \frac{\partial}{\partial p^+}
+ {\zeta} _{\eta}  ^+   {\partial  \over \partial \dot{q}^+ }
+ {\zeta} _{\nu}   ^+  {\partial  \over \partial \dot{p}^+ },
\end{multline}
where
\begin{equation*}
 \xi   =  \xi(t,q,p),
 \qquad
 \eta   =  \eta (t,q,p),
 \qquad
 \nu =  \nu  (t,q,p).
\end{equation*}
The coefficients
\begin{equation*}
{\zeta} _{\eta}   = {\zeta} _{\eta}   ( t, q, p,  \dot{q},  \dot{p} )
=   {\DD}    ( \eta ) -  \dot{q}   {\DD}  ( \xi ),
\qquad
{\zeta} _{\nu}   =  {\zeta} _{\nu}   ( t, q, p,  \dot{q},  \dot{p} )
=   {\DD}    ( \nu  ) -  \dot{p}   {\DD}   ( \xi ),
\end{equation*}
are found according to the standard prolongation
formulas~\cite{bk:Ovsiannikov1978, bk:Ibragimov[1983], bk:Olver[1986]}.
The other coefficients are obtained by the left and right shift operators  $ S_- $ and  $ S_+  $:
\begin{equation*}
\xi   ^- = S _- ( \xi),
\quad
\eta  ^-   = S_-  ( \eta),
\quad
\nu ^-   = S_-  ( \nu ),
\quad
{\zeta} _{\eta}   ^-    = S_-  ( {\zeta} _{\eta}  ),
\quad
{\zeta} _{\nu}   ^-   = S_-  (  {\zeta} _{\nu}  ),
\end{equation*}
\begin{equation*}
\xi   ^+ = S _+ ( \xi),
\quad
\eta  ^+  = S_+ ( \eta),
\quad
\nu ^+   = S_+ ( \nu ),
\quad
{\zeta} _{\eta}    ^+   = S_+  ( {\zeta} _{\eta} ),
\quad
{\zeta} _{\nu}   ^+    = S_+ ( {\zeta} _{\nu}  ).
\end{equation*}
The right and left shift operators  $S_{+}  $ and   $S_{-}  $ are defined in~(\ref{shiftsB}) and~(\ref{shifts}).

Invariance of the functional~(\ref{functional_delay_L}) was analyzed in~\cite{bk:DorodnitsynKozlovMeleshko[2023]}.
Invariance of the functional~(\ref{functional_H_delay}) is developed similarly.
In both Lagrangian and Hamiltonian approaches, the invariance of a delay functional is equivalent to the invariance of the corresponding elementary action.
We go directly to the result.

\begin{theorem}   \label{thm_D1}
 {\bf (Invariance of delay  Hamiltonian)}
The functional~(\ref{functional_H}) is invariant with respect to the group 
transformations  with the generator~(\ref{symmetry_H}) if and only if
\begin{multline}  \label{Group2}
\Omega = X ( \tilde{H} ) + \tilde{H}   D ( \xi )
\\
= \nu^{-}    (\alpha_{1}\dot{q}+\alpha_{2}\dot{q}^{-})
+   p^{-}   (\alpha_{1}D(\eta)+\alpha_{2}D(\eta^{-}))
+  \nu   (\alpha_{3}\dot{q}+\alpha_{4}\dot{q}^{-})
+  p   (\alpha_{3}D(\eta)+\alpha_{4}D(\eta^{-}))
\\
+(\alpha_{2}p^{-}+\alpha_{4}p)   \dot{q}^{-}   D(\xi-\xi^{-})
-\xi\frac{\partial H}{\partial t}
-\eta\frac{\partial H}{\partial q}
-\nu\frac{\partial H}{\partial p}
-\xi^{-}\frac{\partial H}{\partial t^{-}}
-\eta^{-}\frac{\partial H}{\partial q^{-}}
-\nu^{-}\frac{\partial H}{\partial p^{-}}
\\
-HD(\xi)
= 0.
\end{multline}
\end{theorem}

\section{First integrals of delay Hamiltonian equations}
\label{section_first_integrals}

We generalize the form of first integrals for delay ordinary differential equations.
The delay Hamiltonian equations can contain the dependent variables in three points: $t^+$, $t$, and $t^{-}$.
We need to consider two types of conserved quantities: differential first integrals and difference first integrals.

\begin{definition}   \label{differential_first_integral}
Quantity
\begin{equation}   \label{differential_integral}
I(t^{+},t,t^{-},q^{+},q,q^{-},p^{+},p,p^{-})
\end{equation}
is called \textit{a differential first integral} of the delay Hamiltonian equations
if it is constant on the solutions of the delay Hamiltonian equations.
\end{definition}

We require that the differential first integrals~(\ref{differential_integral}) satisfy the equation
\begin{equation}
D(I)=I_{t^{+}}+I_{t}+I_{t^{-}}+I_{q^{+}}\dot{q}^{+}+I_{q}\dot{q}+I_{q^{-}}\dot{q}^{-}+I_{p^{+}}\dot{p}^{+}+I_{\dot{p}}\dot{p}+I_{\dot{p}^{-}}\dot{p}^{-}=0,
\end{equation}
which should hold on the solutions of the considered delay Hamiltonian equations.

In addition to differential first integrals,
we can define the difference ones.

\begin{definition} \label{deference_first_integral}
Quantity
\begin{equation}
J(t,t^{-},q,q^{-},p,p^{-},\dot{q},\dot{q}^{-},\dot{p},\dot{p}^{-})
\end{equation}
is called \textit{a difference first integral} of the delay Hamiltonian
equations if it satisfies the equation
\begin{equation}
(S_{+}-1)J=0
\end{equation}
on the solutions of the delay Hamiltonian equations.
\end{definition}

\section{Hamiltonian identity and Noether-type theorem}
\label{section_Noether}

Noether-type theorems are formulated based on Hamiltonian identity, 
which is  a Hamiltonian version of the Noether operator identity 
used in the Lagrangian approach.

\subsection{Hamiltonian identity}

\begin{lemma}  \label{lem_D3}
{\bf (Hamiltonian identity)}
The following identity holds
\begin{equation}   \label{DelayNoether_b}
\Omega
\equiv
\xi
{ \delta  \tilde{H}  \over \delta t }
+
\eta
{ \delta  \tilde{H}  \over \delta q }
+ \nu
{ \delta  \tilde{H}  \over \delta p }
+  D  ( C )
+ (1 -  S^{+} )    P,
\end{equation}
where
$ \Omega  $   is left hand side of the invariance condition~(\ref{Group2}),
the variations
$ { \delta  \tilde{H}  \over \delta t }  $,
$  { \delta  \tilde{H}  \over \delta q }   $, and
$  { \delta  \tilde{H}  \over \delta p }   $ are given in~(\ref{delay_variation}),
\begin{equation*}
C
=
\eta( \alpha_{4}{p}^{+}
+(\alpha_{2}+\alpha_{3}){p}
+ \alpha_{1}{p}^{-} )
- \xi(
\alpha_{2}  (  p  \dot{q} - p^{-} \dot{q}^{-}  )
+\alpha_{4} ( p^{+}\dot{q} - p \dot{q}^{-}  )
+ H),
\end{equation*}
and
\begin{equation*}
P
=
(\alpha_{2}p^{-}+\alpha_{4}p)D(\eta^{-})
+ \nu^{-}(\alpha_{1}\dot{q}+\alpha_{2}\dot{q}^{-})
- (\alpha_{2}p^{-}+\alpha_{4}p)\dot{q}^{-}D(\xi^{-})
-   \xi^{-}   \frac{\partial H}{\partial t^{-}}
-  \eta^{-} \frac{\partial H}{\partial q^{-}}
-  \nu^{-} \frac{\partial H}{\partial p^{-}}.
\end{equation*}
\end{lemma}

In detail, the identity~(\ref{DelayNoether_b})  takes the form
\begin{multline}   \label{DelayNoether}
\nu^{-}    (\alpha_{1}\dot{q}+\alpha_{2}\dot{q}^{-})
+   p^{-}   (\alpha_{1}D(\eta)+\alpha_{2}D(\eta^{-}))
+  \nu   (\alpha_{3}\dot{q}+\alpha_{4}\dot{q}^{-})
+  p   (\alpha_{3}D(\eta)+\alpha_{4}D(\eta^{-}))
\\
+(\alpha_{2}p^{-}+\alpha_{4}p)   \dot{q}^{-}   D(\xi-\xi^{-})
-\xi\frac{\partial H}{\partial t}
-\eta\frac{\partial H}{\partial q}
-\nu\frac{\partial H}{\partial p}
-\xi^{-}\frac{\partial H}{\partial t^{-}}
-\eta^{-}\frac{\partial H}{\partial q^{-}}
-\nu^{-}\frac{\partial H}{\partial p^{-}}
\\
-HD(\xi)
\equiv
  \xi\left(
    D  [\alpha_{2}(p\dot{q}-p^{-}\dot{q}^{-})+\alpha_{4}(p^{+}\dot{q}-p\dot{q}^{-}) ]
    +D(H)
   - \frac{\partial}{\partial t} (H+H^{+})
   \right)
\\
-\eta \left(
\alpha_{4}\dot{p}^{+}
+(\alpha_{2}+\alpha_{3})\dot{p}
+\alpha_{1}\dot{p}^{-}
+\frac{\partial}{\partial q} (H+H^{+})
\right)
\\
+{\nu}\left(
\alpha_{1}\dot{q}^{+}
+(\alpha_{2}+\alpha_{3})\dot{q}
+\alpha_{4}\dot{q}^{-}
-\frac{\partial}{\partial{p}} (H+H^{+})
\right)
\\
+  D\left[
\eta( \alpha_{4}{p}^{+}
+(\alpha_{2}+\alpha_{3}){p}
+ \alpha_{1}{p}^{-} )
- \xi(
\alpha_{2}  (  p  \dot{q} - p^{-} \dot{q}^{-}  )
+\alpha_{4} ( p^{+}\dot{q} - p \dot{q}^{-}  )
+ H)
\right]
\\
+(1 - S^{+} )   \left[
 (\alpha_{2}p^{-}+\alpha_{4}p)D(\eta^{-})
+ \nu^{-}(\alpha_{1}\dot{q}+\alpha_{2}\dot{q}^{-})
- (\alpha_{2}p^{-}+\alpha_{4}p)\dot{q}^{-}D(\xi^{-})
{\phantom {31 \over 31} }
\right.
 \\
 \left.
-   \xi^{-}   \frac{\partial H}{\partial t^{-}}
-   \eta^{-} \frac{\partial H}{\partial q^{-}}
-   \nu^{-} \frac{\partial H}{\partial p^{-}}
\right].
\end{multline}

\noindent {\it Proof.}
The identity can be verified directly.
\hfill $\Box$

\medskip

\subsection{Noether's theorem for delay Hamiltonian equations}
\label{section_delay_Noether}


Based on the Noether identity,
the most general Noether-type theorem is formulated as follows.

\begin{theorem}    \label{Noether_theorem_H}
{\bf (Noether's theorem)}
Let a Hamiltonian with some given constants $\alpha_{1}$, $\alpha_{2}$, $\alpha_{3}$ and  $\alpha_{4}$
be invariant with respect to a one-parameter group of transformations
with a generator~(\ref{symmetry_H}),
i.e. condition~(\ref{Group2}) hold.
Then,  on the solutions of the local extremal equation
\begin{equation}      \label{quasi_H_b}
F_H
=
\xi
{ \delta  \tilde{H}  \over \delta t }
+
\eta
{ \delta  \tilde{H}  \over \delta q }
+ \nu
{ \delta  \tilde{H}  \over \delta p }
= 0
\end{equation}
there holds the differential-difference relation
\begin{equation}   \label{dd_H}
D  (C)
=  (S^{+}-1)   P.
\end{equation}
\end{theorem}

\noindent {\it Proof.}
The proof follows from the Hamiltonian identity~(\ref{DelayNoether_b}).
\hfill $\Box$

\medskip

For Noether's theorem,
it is convenient to generalize the invariance of a delay  Hamiltonian~(\ref{Group2})
and consider {\it divergence invariance}~\cite{Bess}.

\begin{definition}   \label{divergence_invariance}
The delay Hamiltonian  $ H ( t,t^{-},q,q^{-}, p,  p ^{-})  $
is called  divergence invariant for a generator~(\ref{symmetry_H})
if instead of the condition~(\ref{Group2})
it satisfies the condition
\begin{equation}   \label{dd_invariance}
\Omega  =   {\DD}  ( V  )  +   ( 1  -  S_+  )  W
\end{equation}
with some functions
$
V   (   t  ^+,   t, t ^-,
 q  ^+,  q,  q  ^-,
p  ^+,   p, p ^-     )
$
and
$
W  (     t, t ^-,
      q,  q  ^-,  p,  p ^{-}, 
     \dot{q}, \dot{q}  ^-,
\dot{p}, \dot{p} ^-       )
$
on solutions of local extremal equation~(\ref{quasi_H_b}).
\end{definition}

\begin{corollary}    \label{dd_generalization}
Let delay Hamiltonian functional~(\ref{functional_H_delay}) satisfy
divergence invariance~(\ref{dd_invariance}).
Then,    on solutions of the local extremal  equation~(\ref{quasi_H_b})
there holds   the differential-difference relation
\begin{equation}       \label{dd_divergence}
  {\DD}  ( C - V )
=
( S_+  - 1 )  (  P - W  ).
\end{equation}
\end{corollary}


\subsubsection{Special cases of the Noether  theorem}

Based on the Noether Theorem~\ref{Noether_theorem_H},
it is possible to provide specified results.

\begin{proposition}  \label{proposition_differential}
If the difference in the  differential-difference relation~(\ref{dd_H})
can be presented as a total derivative,
i.e.
\begin{equation}
( S_+ - 1  )  P  = {\DD}  ( V )
\end{equation}
with some function $ V   (   t  ^+,   t, t ^-,
 q  ^+,  q,  q  ^-,
 p  ^+,   p, p  ^-     )
$,
then there holds  the differential first integral
\begin{equation}
I = C - V.
\end{equation}
\end{proposition}

\begin{proposition}  \label{proposition_difference}
If the total derivative  in the  differential-difference relation~(\ref{dd_H})
can be presented as a difference,
i.e.
\begin{equation}
{\DD}  ( C )  =  ( S_+  - 1 )   W,
\end{equation}
with some function
$
W  (     t, t ^-,        q,  q  ^-,    p, p  ^-,
     \dot{q}, \dot{q}  ^-,      \dot{q}, \dot{q}  ^-     )
$,
then, there is the difference  first integral
\begin{equation}
 J =   P  -  W.
\end{equation}
\end{proposition}


It is also possible to consider first integrals,
which hold under some additional conditions.
We present an example of such results.

\begin{proposition}     \label{proposition_difference_constraint}
 {\bf (Local extremal equation with a difference constraint)}
Let a differential-difference relation~(\ref{dd_H})
hold on the solutions of a local extremal equation~(\ref{quasi_H}).
If there is an additional condition
\begin{equation}  \label{constraint_difference}
  ( S_+  - 1 )    P  =0,
\end{equation}
then there is the differential first integral
\begin{equation}
I  = C.
\end{equation}
\end{proposition}

Generally, such additional conditions as~(\ref{constraint_difference})
reduce the solution set of the considered equations.


\subsubsection{Delay canonical Hamiltonian equations}

The system of local extremal equation~(\ref{quasi_H}) and delay parameter equation~(\ref{delay_eq}) is {under}determined.
There are several ways to choose a determined system.
The most important is to consider the delay canonical Hamiltonian equations.

\begin{proposition}
The results of the Theorem~\ref{Noether_theorem_H}  hold for
symmetries~(\ref{symmetry_H}) with   $ \xi   \equiv 0 $
and the delay canonical Hamiltonian equations~(\ref{delay_canonical})
with constant delay~(\ref{delay_eq}).
\end{proposition}

\begin{remark}
It is also possible to consider other determined systems.
For example, one can choose
the symmetries~(\ref{symmetry_H}),(\ref{xi_we_take}) with $ \eta  \equiv 0 $
and the system of equations
\begin{equation}      \label{system_pt}
{ \delta  \tilde{H}  \over \delta p }
= 0,
\qquad
{ \delta  \tilde{H}  \over \delta t }
= 0,
\qquad
 t^+ - t  =  t - t^- = \tau
\end{equation}
or
the symmetries~(\ref{symmetry_H}),(\ref{xi_we_take})     with $ \nu    \equiv 0 $
and the system of equations
\begin{equation}        \label{system_qt}
{ \delta  \tilde{H}  \over \delta q }
= 0,
\qquad
{ \delta  \tilde{H}  \over \delta t }
= 0,
\qquad
 t^+ - t  =  t - t^- = \tau.
\end{equation}
However, the system which consists of all three variational equations  (\ref{delay_variation}) 
and the delay equation~(\ref{delay_eq}) is {over}determined.
\end{remark}

\section{Invariance of the variational  equations}
\label{section_Invariance}

In this section, we consider generators~(\ref{symmetry_H})
with coefficient $ \xi (t,q,p) = \xi (t)  $ given in~(\ref{xi_we_can}), 
i.e., satisfying condition (\ref{xi_condition}). 
The variational equations are assumed to be considered with the delay equation~(\ref{delay_eq}).

\begin{lemma}    \label{three_identities}
The following identities are true for generators~(\ref{symmetry_H}),(\ref{xi_we_can})
\begin{subequations}
\begin{gather}
\label{var_p}
{\delta  \over \delta p}    \Omega
\equiv
X \left(    {\delta  \tilde{H}  \over \delta p} \right)
+ { \partial \eta  \over \partial p }    {\delta \tilde{H}  \over \delta q}
+  \left( { \partial \nu \over \partial p }  +   \dot{\xi}  \right) {\delta  \tilde{H}  \over \delta p},
\\
\label{var_q}
{\delta  \over \delta q}    \Omega
\equiv X \left(    {\delta  \tilde{H}  \over \delta q} \right)
+ \left(   { \partial \eta  \over \partial q }      +  \dot{\xi}  \right)    {\delta  \tilde{H}  \over \delta q}
+   { \partial \nu  \over \partial q }    {\delta  \tilde{H}  \over \delta p},
\\
\label{Var_t}
{\delta  \over \delta t}   \Omega
\equiv X \left(    {\delta  \tilde{H}  \over \delta t} \right)
 +
 2 \dot{\xi}    {\delta  \tilde{H}  \over \delta t}
+   { \partial \eta  \over \partial t }   {\delta  \tilde{H}  \over \delta q}
+   { \partial \nu  \over \partial t }   {\delta  \tilde{H}  \over \delta p},
\end{gather}
\end{subequations}
where $ \Omega $  is given by~(\ref{Group2}) and
$ {\tilde{H}} $  is given by (\ref{H_tilde}).
\end{lemma}

\noindent {\it Proof.}
The identities are verified by direct computation.
\hfill $\Box$

\medskip

\begin{lemma}    \label{one_identity}
The following identity is true for generators~(\ref{symmetry_H}),(\ref{xi_we_can})
\begin{equation}     \label{var_quasi}
\left( \xi {\delta  \over \delta t} + \eta {\delta  \over \delta q}
+ \nu {\delta  \over \delta p} \right) \Omega
\equiv
X \left( \xi
{\delta  \tilde{H}  \over \delta t}
+  \eta {\delta  \tilde{H}  \over \delta q}
+ \nu {\delta  \tilde{H}  \over \delta p} \right)
 +
\dot{\xi} \left( \xi  {\delta  \tilde{H}  \over \delta t}
+  \eta {\delta  \tilde{H}  \over \delta q}
+ \nu {\delta  \tilde{H}  \over \delta p } \right).
\end{equation}
\end{lemma}

\noindent {\it Proof.}
The identity follows from identities of Lemma~\ref{three_identities}.
\hfill $\Box$

\medskip

The identities given above allow us to state the following invariance conditions.

\begin{theorem} \label{invarianceLocalEq}
 The necessary and sufficient condition for the
invariance of the local extremal equation~(\ref{quasi_H_b})
is
\begin{equation}
\left.
\left( \xi {\delta  \over \delta t}
+ \eta {\delta  \over \delta q}
+ \nu {\delta  \over \delta p} \right) \Omega
\right|  _{F_H =0}
=0.
\end{equation}
\end{theorem}

\noindent {\it Proof.}
The results follow from the identity of Lemma~\ref{one_identity}.
\hfill $\Box$

\medskip

\begin{theorem} \label{invarianceofvarEq}
 The necessary and sufficient condition for the
invariance of the delay canonical Hamiltonian equations~(\ref{delay_canonical}) are
\begin{subequations}       \label{conditions_on_Omega}
\begin{gather}
\left.
\frac{\delta  }{\delta p}  \Omega
\right| _{ {\delta  \tilde{H} \over \delta p} =0, \  {\delta  \tilde{H} \over \delta q} =0}
=0,
\\
\left.
\frac{\delta }{\delta q}  \Omega
\right| _{ {\delta  \tilde{H}   \over \delta p} =0, \  {\delta  \tilde{H} \over \delta q} =0}
=0.
\end{gather}
\end{subequations}
\end{theorem}

\noindent {\it Proof.}
We recall that the delay canonical Hamiltonian equations~(\ref{delay_canonical})
can be presented as
\begin{equation*}
{\delta \tilde{H}  \over \delta p} = 0,
\qquad
{\delta \tilde{H}  \over \delta q} = 0.
\end{equation*}
On the solutions of these equations, we rewrite identities~(\ref{var_p})  and~(\ref{var_q})  as
\begin{equation*}  
\begin{array}{l}
{\displaystyle 
\left.
\frac{\delta }{\delta p}  \Omega
\right| _{ {\delta  \tilde{H}  \over \delta p} =0, \  {\delta  \tilde{H}  \over \delta q} =0}
=
\left.
X   \left( {\delta  \tilde{H} \over \delta p} \right)
\right| _{ {\delta  \tilde{H}  \over \delta p} =0, \  {\delta  \tilde{H}  \over \delta q} =0},
}
\\
\\
{\displaystyle 
\left.
\frac{\delta }{\delta q}  \Omega
\right| _{ {\delta  \tilde{H}  \over \delta p} =0, \  {\delta  \tilde{H}  \over \delta q} =0}
=
\left.
X \left(   {\delta \tilde{H}  \over \delta q}   \right)
\right| _{ {\delta  \tilde{H}  \over \delta p} =0, \  {\delta  \tilde{H}  \over \delta q} =0}.
}
\end{array}
\end{equation*}
These equations state that conditions~(\ref{conditions_on_Omega}) are equivalent to the invariance of the delay canonical Hamiltonian equations.
\hfill $\Box$

\medskip

\begin{remark}
The statement of  Theorem~\ref{invarianceofvarEq}
also holds for divergence invariant Hamiltonians
~(\ref{dd_invariance})
because the variational operators
$  {\delta  \over \delta p}  $
and
$  {\delta  \over \delta q} $
annihilate the differential and difference divergence terms.
Here, the variational operator~(\ref{varoperator1_delay_p})  and~(\ref{varoperator1_delay_q})
 are  extended to the third point  $ t^+$  as
\begin{equation*}
\begin{array}{l}
{\displaystyle 
\frac{\delta}{\delta p}
=
\frac{\partial}{\partial p}
- D\frac{\partial}{\partial\dot{p}}
+ S_+ \left(
\frac{\partial}{\partial p^- }
- D\frac{\partial}{\partial\dot{p}^- }
\right)
+ S_- \left(
\frac{\partial}{\partial p^+ }
- D\frac{\partial}{\partial\dot{p}^+ }
\right),
}
\\
\\
{\displaystyle 
\frac{\delta}{\delta q}
=
\frac{\partial}{\partial q}
- D \frac{\partial}{\partial\dot{q}}
+ S_+ \left(
\frac{\partial}{\partial q ^- }
- D \frac{\partial}{\partial \dot{q} ^- }
\right)
+ S_- \left(
\frac{\partial}{\partial q ^+ }
- D \frac{\partial}{\partial \dot{q} ^+ }
\right).
}
\end{array}
\end{equation*}
\end{remark}

\section{Examples}
\label{Section_Examples}

We continue with Examples~\ref{Non_degenerate case} and~\ref{Degenerate case}.
Now, we examine the delay canonical Hamiltonian equations for invariance and conserved quantities.


\begin{example}
{\bf Delay {non}degenerate harmonic oscillator
(continuation of Example~\ref{Non_degenerate case})}

In Example~\ref{Non_degenerate case},
we derived the delay Hamiltonian
\begin{equation}
H =  pp^{-}+qq^{-},
\end{equation}
which provides the delay canonical Hamiltonian equations
\begin{subequations}   \label{HamEq32}
\begin{gather}
\dot{q}^{+}+\dot{q}^{-}
= p^{+}+p^{-},
\\
\dot{p}^{+}+\dot{p}^{-}
=- q^{+}  -  q^{-}.
\end{gather}
\end{subequations}
We consider these equations with delay equation~(\ref{delay_eq}).
For the delay Legendre transformation, there were used coefficients
\begin{equation*}
\alpha_{1}= 1,
\qquad
\alpha_{2}=0,
\qquad
\alpha_{3}=0,
\qquad
\alpha_{4}=1.
\end{equation*}

Delay equations~(\ref{HamEq32}) admit  symmetry operators
\begin{multline}
X_{1}=\sin t {\frac{\partial}{\partial q}} + \cos t {\frac{\partial}{\partial p}},
\qquad
X_{2}=\cos t {\frac{\partial}{\partial q}}-\sin t {\frac{\partial}{\partial p}},
\\
X_{3}={\frac{\partial}{\partial t}},
\qquad
X_{4}=q{\frac{\partial}{\partial q}}+p{\frac{\partial}{\partial p}},
\qquad
X_{5}=p{\frac{\partial}{\partial q}}-q{\frac{\partial}{\partial p}}.
\end{multline}
We note that
symmetry  $X_{5}$ has no corresponding Lie point symmetry in the Lagrangian approach.


For symmetry $X_{1}$, the Hamiltonian is divergence invariant
\begin{equation*}
\Omega  _1
=D ( \cos t^{-} q   +   \cos t \  q^{-}    ).
\end{equation*}
The symmetry gives the differential-difference relation
\begin{equation*}
D (  \sin t \  ( p^+ + p^- )   -   \cos t^{-} q -   \cos t  \  q^{-} )
= ( S_+  - 1 )  (  \cos t^{-}  \dot{q}    -   \sin t^{-} q ).
\end{equation*}
Since the right-hand side of the differential-difference relation can be rewritten as a total derivative
\begin{equation*}
( S_+  - 1 )  (  \cos t^{-}  \dot{q}    -   \sin t^{-} q )
= D (   cos t \  {q^+}   - cos t^{-}  {q} ),
\end{equation*}
we obtain the differential integral
\begin{equation}
I_1
=
 \sin t  \    ({p}^{+}+{p}^{-})
-    \cos t  \   (q^{+}+q^{-}) .
\end{equation}


Symmetry  $X_{2}$ also satisfies divergence invariance of the Hamiltonian
\begin{equation*}
\Omega _2
= D(  -   \sin t^{-}  q    -    \sin t  \    q^{-}  ).
\end{equation*}
We obtain the  differential-difference relation
\begin{equation*}
D (  \cos t \  ( p^+ + p^- )   +    \sin  t^{-} q  +    \sin  t  \  q^{-} )
= ( S_+  - 1 )  (   - \sin t^{-}  \dot{q}    -   \cos  t^{-} q ).
\end{equation*}
Rewriting  the right-hand side of the  differential-difference relation as the total derivative
\begin{equation*}
 ( S_+  - 1 )  (   - \sin t^{-}  \dot{q}    -   \cos  t^{-} q )
= D ( - \sin  t  \ q^+  +  \sin  t^- q  ),
\end{equation*}
we obtain the  differential integral
\begin{equation}
I_2 =   \cos t   \     ({p}^{+}+{p}^{-})    +    \sin t    \
(q^{+}+q^{-}) .
\end{equation}


Thought symmetry $X_{3}$  is a variational symmetry of the Hamiltonian,
i.e.
\begin{equation*}
\Omega _3 =  0,
\end{equation*}
It needs other equation than the canonical Hamiltonian equation~(\ref{HamEq32}),(\ref{delay_eq})
for deriving the differential-difference relation.
For example, one can consider systems of equations~(\ref{system_pt}) and~(\ref{system_qt}).
We will not consider this symmetry since we are interested in the delay canonical Hamiltonian equations.


For symmetry  $X_{4}$, we find
\begin{equation*}
\Omega _4 = 2 (   p^-  \dot{q} +   p  \dot{q} ^-   - p p^- - q q^- ).
\end{equation*}
The symmetry is neither variational symmetry nor divergence symmetry of the Hamiltonian.
Thus, there is no differential-difference relation.

Despite symmetry $X_{4}$ is not a symmetry for the delay Hamiltonian,
it is a symmetry of the delay canonical Hamiltonian equations.
Invariance of the  Hamiltonian equations  follows from
\begin{equation*}
{\delta \over \delta p} \Omega _4
=
{\delta \tilde{H}  \over \delta p},
\qquad
{\delta \over \delta q} \Omega _4
=
{\delta \tilde{H}  \over \delta q}
\end{equation*}
that makes the conditions of Theorem~\ref{invarianceofvarEq} satisfied.


Symmetry  $X_{5}$ is a divergence symmetry of the Hamiltonian
\begin{equation*}
\Omega _5 =  D (  - p p^-  - q q^-  )  = D ( - H ).
\end{equation*}
it provides the differential-difference relation
\begin{equation*}
D (   p  \  ( p^+ + p^- )    +  p p^-   +  q q^-   )
= ( S_+  - 1 )  (      p   \dot{p} ^- - q^- \dot{q} -  q  p ^-   +  q^- p   ).
\end{equation*}
Converting this relation to differential or difference first integral is not possible.

The differential first integrals  $I_1  $ and $ I_2 $ can be used to find solutions to the delay Hamiltonian equations~(\ref{HamEq32}). In fact, setting the values of these first integrals as
\begin{equation*}
I_1 = A, 
\qquad 
I_2 = B,
\end{equation*}
where $ A $ and $ B $ are constant, we obtain the following solution 
\begin{subequations}   \label{sol23}
\begin{gather}
q^{+}+q^{-}= - A \cos t  + B \sin t, \\
{p}^{+}+{p}^{-}=  A\sin t + B  \cos t.
\end{gather}
\end{subequations}

Using this representation, one can construct a solution to a Cauchy problem with the initial values
\begin{equation*}
q(t) = \varphi(t), 
\qquad p(t) = \psi(t), 
\qquad 
t \in [-2\tau, 0],
\end{equation*}
where the functions $ \varphi(t) $ and $ \psi(t) $ are assumed to be continuous. 
By rewriting relations (\ref{sol23}) in shifted form 
\begin{subequations}   \label{recursion}
\begin{gather}
q(t) + q(t - 2\tau) = - A \cos(t - \tau) + B \sin(t - \tau), \\
p(t) + p(t - 2\tau) = A \sin(t - \tau) + B \cos(t - \tau),
\end{gather}
\end{subequations}
we find the constants $ A $ and $ B $ as
\begin{equation*}
A = - \cos(\tau) \left(\varphi(0) + \varphi(-2\tau)\right) - \sin(\tau) \left(\psi(0) + \psi(-2\tau)\right),
\end{equation*}
\begin{equation*}
B = - \sin(\tau) \left(\varphi(0) + \varphi(-2\tau)\right) + \cos(\tau) \left(\psi(0) + \psi(-2\tau)\right).
\end{equation*}

Applying relations~(\ref{recursion}), we derive the solution of the Cauchy problem for $ t \in [0, 2\tau] $
\begin{equation*}
q(t) = - \varphi(t - 2\tau) - A \cos(t - \tau) + B \sin(t - \tau),
\end{equation*}
\begin{equation*}
p(t) = - \psi(t - 2\tau) + A \sin(t - \tau) + B \cos(t - \tau).
\end{equation*}
Then, for $ t \in [2\tau, 4\tau] $
\begin{equation*}
q(t) = \varphi(t - 4\tau) + A \cos(t - 3\tau) - B \sin(t - 3\tau) - A \cos(t - \tau) + B \sin(t - \tau),
\end{equation*}
\begin{equation*}
p(t) = \psi(t - 4\tau) - A \sin(t - 3\tau) - B \cos(t - 3\tau) + A \sin(t - \tau) + B \cos(t - \tau).
\end{equation*}
This process can be continued recursively. 
Using these relations, one can find the solution $ (q(t), p(t)) $ for $ t \in [0, \infty) $. 
Unlike the standard method of solving DODEs, 
namely the method of steps~\cite{bk:Elsgolts[1955]}, 
which requires integration, the recursive procedure outlined above does not involve any integration.

\end{example}


\begin{example}
{\bf  Delay degenerate harmonic oscillator (continuation of   Example~\ref{Degenerate case}) }

We continue Example~(\ref{Degenerate case}) with $ \phi ( q , q^{-})  =  { 1 \over 2 }  (q+q^{-})^{2} $ 
and consider the delay Hamiltonian
\begin{equation}
H=  { 1 \over 2 }  (p+p^{-})^{2}
+   { 1 \over 2 }  (q+q^{-})^{2},
\end{equation}
which provides   the delay canonical Hamiltonian equations
\begin{subequations}    \label{example_2_deg}
\begin{gather}
\dot{q}^{+}+2\dot{q}+\dot{q}^{-}
=p^{+}+2p+p^{-},
\\
\dot{p}^{+}+2\dot{p}+\dot{p}^{-}
= - q^{+} -  2q -  q^{-}.
\end{gather}
\end{subequations}
There were used coefficients
\begin{equation*}
\alpha_{1}=\alpha_{2}=\alpha_{3}=\alpha_{4}=1.
\end{equation*}

The system of delay equations~(\ref{example_2_deg}) admits the symmetry operators
\begin{multline}
X_{1}=\sin t {\frac{\partial}{\partial q}} + \cos t {\frac{\partial}{\partial p}},
\qquad
X_{2}=\cos t {\frac{\partial}{\partial q}} - \sin t  {\frac{\partial}{\partial p}},
\\
X_{3}={\frac{\partial}{\partial t}},
\qquad
X_{4}=q{\frac{\partial}{\partial q}}+p{\frac{\partial}{\partial p}},
\qquad
X_{5}=p{\frac{\partial}{\partial q}}-q{\frac{\partial}{\partial p}}.
\end{multline}


Symmetry  $X_{1}$  is a divergence symmetry of the delay Hamiltonian
\begin{equation*}
\Omega_1
=D(  ( \cos t+\cos t^{-} )   ({q+q^{-}}) ).
\end{equation*}
The symmetry  yields the following differential-difference relation
\begin{multline*}
D (   \sin t  \  ( p^+ + 2 p +  p^- )    -     ( \cos t+\cos t^{-} )   ({q+q^{-}})   )
\\
= ( S_+  - 1 )  (      \cos t^{-}    (  \dot{q} +  \dot{q} ^- )
-    \sin t^{-}    ( {q} +  {q} ^- )  ).
\end{multline*}
Using the presentation of the right-hand side as a total derivative
\begin{equation*}
 ( S_+  - 1 )  (      \cos t^{-}    (  \dot{q} +  \dot{q} ^- )
-    \sin t^{-}    ( {q} +  {q} ^- )  )
= D   (  \cos t \     (  {q}^+  +  {q} )    -  \cos t^{-}    (  {q} +  {q} ^- )  ),
\end{equation*}
we obtain the  first differential integral
\begin{equation}
I_1 =  \sin t  \  ({p}^{+}+2p+{p}^{-})   - \cos t  \  ({q^{+}+2q+q^{-}}).
\end{equation}


Operator $X_{2}$ also provides a divergence symmetry of the delay Hamiltonian
\begin{equation*}
\Omega_2
=  D(  -  (\sin t+\sin t^{-})   ({q+q^{-}}) ).
\end{equation*}
It yields the differential-difference relations
\begin{multline*}
D (   \cos t  \  ( p^+ + 2 p +  p^- )    +     ( \sin t+\sin t^{-} )   ({q+q^{-}})   )
\\
= ( S_+  - 1 )
(      - \sin  t^{-}    (  \dot{q} +  \dot{q} ^- )
-    \cos  t^{-}    ( {q} +  {q} ^- )  ).
\end{multline*}
Rewriting the right-hand side as a total derivative
\begin{equation*}
 ( S_+  - 1 )
 (      - \sin  t^{-}    (  \dot{q} +  \dot{q} ^- )
-    \cos  t^{-}    ( {q} +  {q} ^- )  )
= D   (  -  \sin t \     (  {q}^+  +  {q} )     +   \sin  t^{-}    (  {q} +  {q} ^- )  ),
\end{equation*}
we get the  first differential integral
\begin{equation}
I_2 =   \cos t  \   ({p}^{+}+2p+{p}^{-})
+  \sin t  \    ({q^{+}+2q+q^{-}}).
\end{equation}

Setting the first integrals $I_1 $ and $I_2$ equal to constants $A$ and $B$,
we obtain the relations
\begin{subequations}
\begin{gather}
q^{+}+2q+q^{-}=  -A\cos t   + B\sin t,
\\
{p}^{+}+2p+{p}^{-} =A\sin t + B\cos t.
\end{gather}
\end{subequations}
These relations can be used for the recursive presentation of the solution
similar to that given in the previous example.

\end{example}

\section{Concluding remarks}
\label{section_conclusion}

The paper develops a Hamiltonian formalism for delay ordinary differential equations. 
The delay canonical Hamiltonian equations, 
provided by delay Hamiltonians, correspond to variational DODEs (the Elsgolts equations) 
for first-order Lagrangians with one delay. 
We consider delay Lagrangians, which are quadratic in the derivatives. 
The relation between delay Hamiltonians 
and delay Lagrangians is given by the delay analog of the Legendre transformation (\ref{delay_Legendre}) 
and compatibility condition  (\ref{requirement_p}).

The recent article~\cite{bk:DorodnitsynKozlovMeleshko[2023]} devoted
to delay variational equations in the Lagrangian framework. 
In particular, it was shown that for the invariance of DODEs, one needs
to require the DODE itself and the invariance of a delay parameter equation.
 The invariance of the delay parameter restricts group transformations of the independent variable. 
These results are relevant to DODEs in the Hamiltonian approach.

This article presents criteria for the invariance of delay
functionals in the Hamiltonian framework (called invariance of delay Hamiltonians). 
The Hamiltonian operator identity, which relates invariant Hamiltonians, 
corresponding variational equations, and conserved quantities, was developed. 
On the basis of this identity, Noether's theorem is formulated. 
It allows us to find conserved quantities for delay variational equations with symmetry. 
If there are sufficiently many first integrals, 
they can be used to provide a recursive presentation of the solutions for the Cauchy problem. 
In the general case, the theorem is stated for an {under}determined system, 
which consists of the local extremal equation provided with regular delay parameter. 
This general formulation is used to specify Noether's theorem to particular determined systems. 
The main choice is the system of the delay canonical Hamiltonian equations. 
Several examples illustrated the obtained results. 
The relation between the invariance of the delay variational equations and the Hamiltonian invariance is also provided. 
The necessary and sufficient conditions for the invariance of the local extremal equation 
and invariance of the delay canonical Hamiltonian equations were established. 
The results of the paper can be generalized to vector-valued dependent variables.


\section*{Appendices}

\appendix

\section{Comment on derivation of Hamiltonian equations}
\label{Comment_derivation}

We provide a classical derivation of the canonical  Hamiltonian equations in detail for completeness. 
Let $ L(t, q, \dot{q}) $ be a nondegenerate Lagrangian,
meaning that $ \frac{\partial^{2} L}{\partial \dot{q}^{2}} \neq 0 $.
Consider the Euler-Lagrange equation
\begin{equation}\label{eq:aug16.1}
\frac{\delta L}{\delta q}
= \frac{\partial L}{\partial q} - D\left(\frac{\partial L}{\partial \dot{q}}\right)
= 0.
\end{equation}

To construct the Hamiltonian and derive the canonical Hamiltonian equations corresponding to the Euler-Lagrange equation,
one typically follows the following approach~\cite{bk:Ostrogradsky}. 
Introducing the variable $ r = \dot{q} $, the Lagrangian function becomes
\begin{equation*}
\tilde{L}(t, q, r) = L(t, q, r).
\end{equation*}
Let 
\begin{equation*}
p = \frac{\partial \tilde{L}}{\partial r}(t, q, r) . 
\end{equation*}
As the Lagrangian $ L $ is nondegenerate,
using the inverse function theorem,
one can find $ r = f(t, q, p) $. From this definition of the function $ f $, one has
\begin{equation*}
\frac{\partial f}{\partial p}
= \left(\frac{\partial^{2} \tilde{L}}{\partial r^{2}}\right)^{-1},
\qquad
\frac{\partial f}{\partial q}
= -  \left(\frac{\partial^{2} \tilde{L}}{\partial r^{2}}\right)^{-1} 
\frac{\partial^{2} \tilde{L}}{\partial r \partial q} . 
\end{equation*}
The Euler-Lagrange equation~(\ref{eq:aug16.1}) reduces to the equations
\begin{equation}\label{eq:aug16.2}
\dot{q} = f,
\qquad
\dot{p} = \frac{\partial \tilde{L}}{\partial q}.
\end{equation}

The Legendre transformation
\begin{equation}  \label{Legendre_app}
H = \dot{q} \frac{\partial L}{\partial \dot{q}} - L
= r \frac{\partial \tilde{L}}{\partial r} - \tilde{L}
\end{equation}
defines the Hamiltonian function:
\begin{equation}  \label{Legendre2}
H(t, q, p )
= p f(t, q, p) - \hat{L}(t, q, p),
\end{equation}
where $ \hat{L}(t, q, p) = \tilde{L}(t, q, f(t, q, p)) $. 
Calculating the variational derivatives of the function $ H(t, q, p) $
\begin{equation*}
 \frac{\partial H}{\partial p}
= f + p \frac{\partial f}{\partial p} - \frac{\partial \hat{L}}{\partial p}
= f + p \frac{\partial f}{\partial p} - \frac{\partial \tilde{L}}{\partial r} \frac{\partial f}{\partial p}
= f,
\end{equation*}
\begin{equation*}
 \frac{\partial H}{\partial q}
= p \frac{\partial f}{\partial q} - \frac{\partial \hat{L}}{\partial q}
= p \frac{\partial f}{\partial q} - \frac{\partial \tilde{L}}{\partial q}
- \frac{\partial \tilde{L}}{\partial r} \frac{\partial f}{\partial q}
= p \frac{\partial f}{\partial q} - \frac{\partial \tilde{L}}{\partial q} - p \frac{\partial f}{\partial q}
= -\frac{\partial \tilde{L}}{\partial q}.
\end{equation*}
one concludes that equations~(\ref{eq:aug16.2}) can be rewritten in the form:
\begin{equation*}
\dot{q} = \frac{\partial  H}{\partial  p}, 
\qquad 
\dot{p} = -\frac{\partial H}{\partial  q}.
\end{equation*}

\section{The algorithm to construct the Legendre transformation for an extended case}
\label{section_algorithm}

Here, we show that the approach developed in sections \ref{section_delay_ODEs} 
and~\ref{section_Hamiltonian_formalism} can be extended to more general Lagrangians. 
It is derived that we can apply this approach to Lagrangians 
which is quadratic with respect to $q$ and $\dot{q}^{-}$
\begin{equation}   \label{eq:gen_Lagr}
{L}  
=
{ \alpha \over 2 } {\dot{q}}^{2} 
+  \beta \dot{q} \dot{q}^{-}
+ { \gamma \over 2 } (\dot{q}^{-})^{2}
+h_{1}\dot{q}
+h_{2}\dot{q} ^{-}  
-  \phi, 
\qquad 
\alpha  \gamma - \beta ^2  \neq  0,  
\end{equation}
where $\alpha(q,q^{-})$, $\beta(q,q^{-})$, and $\gamma(q,q^{-})$ are
functions of $q$ and $q^{-}$ in contrast to the previous study. 
The functions $h_{1} (q,q^{-})   $ and  $h_{1} (q,q^{-})   $ have special forms 
(see (\ref{eq:mar21.10_mod})) and the potential $\phi  (q,q^{-}) $ remains as before.

We start with the general form of the time-independent delay Lagrangian
\begin{equation}    \label{Lagrangian_B} 
L = L(q,q^{-}, \dot{q} , \dot{q}^{-}). 
\end{equation} 
{Let $ r $ and  $ r^{-} $  be new dependent variables, 
and define the function $ \tilde{L}(q, q^{-}, r, r^{-}) = L(q, q^{-}, r, r^{-}) $. 
Furthermore, if there is no ambiguity, the function $ \tilde{L}(q, q^{-}, r, r^{-}) $ 
will simply be written as $ L(q, q^{-}, r, r^{-}) $.
}

The Lagrangian (\ref{Lagrangian_B}) provides the Hamiltonian function 
\begin{equation}   \label{eq:apr3.4}
H(q, q ^{-}, p, p ^{-} )
=p^{-}(\alpha_{1}r+\alpha_{2}r^{-})
+p(\alpha_{3}r+\alpha_{4}r^{-})
-L(q,q^{-},r,r^{-}),
\end{equation}
with 
\begin{equation*}
r=g_{1}(q,q^{-},p,p^{-}),
\qquad
r^{-}=g_{2}(q,q^{-},p,p^{-})
\end{equation*}
found from the relations
\begin{equation}  \label{eq:mar21.1}
\alpha_{1}p^{-}+\alpha_{3}p=  { \partial  L \over \partial r},
\qquad
\alpha_{2}p^{-}+\alpha_{4}p=  { \partial  L \over \partial r^{-}}.
\end{equation}
Because $r$ and $r ^{-}$ playing the role of $\dot{q}$ and
$\dot{q}^{-} $, then one needs to require that $r=S_{+} (r^{-}) $, 
which means that
\begin{equation*}
g_{1}(q,q^{-},p,p^{-})=g_{2}(q^{+},q,p^{+},p).
\end{equation*}
{Here, $ \alpha _i (q,q^{-}) $, $ i =  1,2,3,4$, are function to be found.} 
It is assumed  that relations~(\ref{eq:mar21.1}) can be solved with respect to $p$ and $p^{-} $, 
which is guaranteed by the condition
\begin{equation}   \label{eq:mar16.5-1}
\Delta=  \alpha_{1}\alpha_{4} -   \alpha_{2}\alpha_{3} \neq 0.
\end{equation}

The delay canonical Hamiltonian equations are obtained by applying 
the variational operators to the function
\begin{equation*}
\tilde{H} 
=p^{-}(\alpha_{1}\dot{q}+\alpha_{2}\dot{q}^{-})
+p(\alpha_{3}\dot{q}+\alpha_{4}\dot{q}^{-})
-H. 
\end{equation*}
These equations
\begin{equation*}
\frac{\delta \tilde{H}}{\delta p}=0,
\qquad
\frac{\delta \tilde{H}}{\delta q}=0
\end{equation*} 
take the form 
\begin{subequations}  
\begin{gather}  
\label{eq:el_pha}
\alpha _1 ^+    \dot{q} ^+ 
+ (   \alpha _2 ^+   +   \alpha _3 )  \dot{q}
+ \alpha _4    \dot{q} ^-   
=  
H_{p} +H_{p}^{+}, 
\\
\label{eq:el_qha}
\begin{array}{l}
{ \displaystyle 
D ( 
\alpha _4 ^+    p ^+ 
+ (   \alpha _2 ^+   +   \alpha _3 )  p
+ \alpha _1    p ^- 
) 
= 
-  H_{q}  -  H_{q}^{+}  
+ p ^{+}
\left(
 { \partial \alpha_{3} ^{+} \over \partial  q }   \dot{q} ^{+}
+ { \partial \alpha_{4}  ^{+} \over \partial  q }   \dot{q}
\right)
}
\\
{ \displaystyle 
+ 
p
\left(   { \partial \alpha_{1} ^{+} \over \partial  q } \dot{q}  ^{+}
+  { \partial \alpha_{2} ^{+}  \over \partial  q }   \dot{q}
+  { \partial \alpha_{3}  \over \partial  q }   \dot{q}
+ { \partial \alpha_{4}  \over \partial  q }   \dot{q}^{-}
\right)
+ 
p^{-}
\left(   { \partial \alpha_{1}  \over \partial  q } \dot{q}  
+  { \partial \alpha_{2}  \over \partial  q }   \dot{q}^{-}
\right)
}. 
\end{array}
\end{gather}
\end{subequations}

\begin{remark}
Equation  (\ref{eq:el_qha}) can also be presented as  
\begin{multline*}
{ \displaystyle 
\alpha _4 ^+    \dot{q} ^+ 
+ (   \alpha _2 ^+   +   \alpha _3 )  \dot{q}
+ \alpha _1    \dot{q} ^- 
) 
= 
-  H_{q}  -  H_{q}^{+}  
+ 
\left(
 { \partial \alpha_{3} ^{+} \over \partial  q }  
- { \partial \alpha_{4}  ^{+} \over \partial  q  ^{+}}  
\right)    p ^{+}  \dot{q} ^{+}
}
\\
{ \displaystyle 
+ 
\left(  { \partial \alpha_{1} ^{+} \over \partial  q }
-   { \partial \alpha_{2} ^{+}  \over \partial  q ^{+} }  
\right)  p \dot{q} ^{+}
+ 
\left( 
-   { \partial \alpha_{3}  \over \partial  q^{-} }  
+ { \partial \alpha_{4}  \over \partial  q }   
\right) p \dot{q}^{-}
+ 
\left(   -  { \partial \alpha_{1}  \over \partial  q ^{-}  } 
+  { \partial \alpha_{2}  \over \partial  q }  
\right)   p^{-}  \dot{q}^{-}  
}. 
\end{multline*}
\end{remark}


Solving~(\ref{eq:mar21.1}) with respect to $p^{-}$ and $p$, one
finds
\begin{equation}   \label{eq:gen-1}
p=\frac{ \displaystyle  \alpha_{1}\frac{\partial L}{\partial{r}^{-}}  - \alpha_{2}\frac{\partial L}{\partial r} }
{ \alpha_{1}\alpha_{4} - \alpha_{2}\alpha_{3}},
\qquad 
p^{-}
=\frac{  \displaystyle  \alpha_{4}\frac{\partial L}{\partial r} - \alpha_{3}\frac{\partial L}{\partial{r}^{-}}}
{\alpha_{1}\alpha_{4}-\alpha_{2}\alpha_{3}},
\end{equation}
The compatibility condition $p=S^{+}(p^{-})$ leads to the relation
\begin{equation}   \label{equt1-1}
\frac{1}{\Delta}
\left(\alpha_{2}\frac{\partial L}{\partial r}-\alpha_{1}\frac{\partial L}{\partial{r}^{-}}\right)
+ \frac{1}{\Delta^{+}}
\left(\alpha_{4}^{+}\frac{\partial L^{+}}{\partial r^{+}}-\alpha_{3}^{+}\frac{\partial L^{+}}{\partial r}\right)
=0,
\end{equation}
which should be equal identically to zero.

Differentiating equation~(\ref{equt1-1}) with respect to $r^{+}$ and
$r^{-}$, one derives
\begin{equation*}
\alpha_{4}^{+}\frac{\partial^{2}L^{+}}{\partial (r^{+}) ^{2}}
-\alpha_{3}^{+}\frac{\partial^{2}L^{+}}{\partial r\partial r^{+}}=0,
\qquad
\alpha_{2}\frac{\partial^{2}L}{\partial r\partial r^{-}}
-\alpha_{1}\frac{\partial^{2}L}{\partial (r^{-})^{2}}=0.
\end{equation*}
Applying the backward shift operator $S_{-}$ to the first equation, the latter
system of equations can be rewritten as
\begin{equation}     \label{eq:mar18.1-1}
\alpha_{4}\frac{\partial^{2}L}{\partial r\,^{2}}
-\alpha_{3}\frac{\partial^{2}L}{\partial r \partial r ^{-} }=0,
\qquad
\alpha_{2}\frac{\partial^{2}L}{\partial r\partial r^{-}}
-\alpha_{1}\frac{\partial^{2}L}{\partial (r^{-})^{2}}=0.
\end{equation}

Assuming that $\frac{\partial^{2}L}{\partial r\partial r^{-}}\neq0$,
equations~(\ref{eq:mar18.1-1}) give
\begin{equation}   \label{eq:mar21.2}
\alpha_{3} 
= \frac{ {\displaystyle  \frac{\partial^{2}L}{\partial r ^{2}} }}{{\displaystyle\frac{\partial^{2}L}{\partial r \partial r ^{-} }}}
\alpha_{4},
\qquad
\alpha_{2} =
\frac{ {\displaystyle \frac{\partial^{2}L}{\partial (r^{-})^{2}}}}{{\displaystyle\frac{\partial^{2}L}{\partial r\partial r^{-}}}}
\alpha_{1}.
\end{equation}
Notice that then
\begin{equation}   \label{eq:mar16.5-1_mod}
\Delta
=
\left(
1 - 
\frac{\displaystyle \frac{\partial^{2}L}{\partial r^{2}}}
{\displaystyle \frac{\partial^{2}L}{\partial  r  \partial r ^{-} }}
\frac{\displaystyle \frac{\partial^{2}L}{\partial (r^{-}) ^{2}}}
{\displaystyle \frac{ \partial^{2}L}{\partial r\partial r^{-}}} 
\right)\alpha_{1}\alpha_{4}
\neq0.
\end{equation}
As $\alpha_{i}$, $i=1,2,3,4 $,  are functions of $q$ and $q^{-}$,
relations~(\ref{eq:mar21.2}) provide that there exist functions
$k_{1}(q,q^{-})$ and $k_{2}(q,q^{-})$ such that
\begin{equation}    \label{eq:mar16.6-1}
\frac{\partial^{2}L}{\partial r^{2}}-k_{1}
\frac{\partial^{2}L}{\partial r\partial r^{-}}=0,
\qquad
\frac{\partial^{2}L}{\partial (r^{-})^{2}}-k_{2}
\frac{\partial^{2}L}{\partial r\partial r^{-}}
=0.
\end{equation}

By virtue of~(\ref{eq:mar16.5-1_mod}), one has $(1 - k_{1}k_{2} ) \alpha_{1} \alpha_{4} \neq 0 $. 
Integrating~(\ref{eq:mar16.6-1}), one gets
\begin{equation*}   
\frac{\partial L}{\partial r}-k_{1}\frac{\partial L}{\partial r^{-}}=f_{1}(q,q^{-},r^{-}),
\qquad
\frac{\partial L}{\partial r^{-}}-k_{2}\frac{\partial L}{\partial r}=-f_{2}(q,q^{-},r).
\end{equation*}
which can be rewritten as
\begin{equation}   \label{eq:mar16.9-1}
\begin{array}{c}
{\displaystyle
(1-k_{1}k_{2})\frac{\partial L}{\partial r}=f_{1}(q^{-},q,r^{-})-k_{1}f_{2}(q,q^{-},r)},\\
\\
{\displaystyle
(1-k_{1}k_{2})\frac{\partial L}{\partial r^{-}}=k_{2}f_{1}(q^{-},q,r^{-})-f_{2}(q,q^{-},r)}.
\end{array}
\end{equation}

Taking the mixed derivative of $L$, we obtain 
\begin{equation}   \label{eq:mar16.10-1}
\frac{\partial f_{1}}{\partial r^{-}}(q,q^{-},r^{-})
=-\frac{\partial f_{2}}{\partial r}(q,q^{-},r).
\end{equation}
Differentiating latter with respect to $r$ and $r^{-}$, one has
\begin{equation*}
\frac{\partial^{2}f_{2}}{\partial r^{2}}(q,q^{-},r)=0, 
\qquad 
\frac{\partial^{2}f_{1}}{\partial (r^{-})^{2}}(q,q^{-},r^{-})=0. 
\end{equation*}
Therefore,
\begin{equation*}
\begin{array}{c}
f_{1}(q,q^{-},r^{-})=\varphi_{1}(q,q^{-})r^{-}+\tilde{\psi}_{1}(q,q^{-}),\\
f_{2}(q,q^{-},r)=\varphi_{2}(q,q^{-})r+\tilde{\psi}_{2}(q,q^{-}),
\end{array}
\end{equation*}
where $\varphi_{i}$ and $\tilde{\psi}_{i}$, $i=1,2$, are arbitrary
functions. Equation~(\ref{eq:mar16.10-1}) gives
\begin{equation*}
\varphi_{2}(q,q^{-})=-\varphi_{1}(q,q^{-}).
\end{equation*}
Substituting these functions into~(\ref{eq:mar16.9-1}), one obtains
\begin{equation*}   
\begin{array}{c}
{\displaystyle
(1-k_{1}k_{2})\frac{\partial L}{\partial r}
=\varphi_{1}(q,q^{-})r^{-}+\tilde{\psi}_{1}(q,q^{-})
-k_{1}( - \varphi_{1}(q,q^{-})r+\tilde{\psi}_{2}(q,q^{-}))},\\
\\
{\displaystyle
(1-k_{1}k_{2})\frac{\partial L}{\partial r^{-}}
=k_{2}(\varphi_{1}(q,q^{-})r^{-}+\tilde{\psi}_{1}(q,q^{-}))
- ( - \varphi_{1}(q,q^{-})r+\tilde{\psi}_{2}(q,q^{-}))}.
\end{array}
\end{equation*}

Integrating the first equation, one finds
\begin{equation*}
(1-k_{1}k_{2})L
=\varphi_{1}(q,q^{-}) r  r^{-} 
+\tilde{\psi}_{1}(q,q^{-})r 
-k_{1} 
\left( 
-\frac{1}{2}\varphi_{1}(q,q^{-})r^{2}+r\tilde{\psi}_{2}(q,q^{-})
\right)
+\chi  (q,q^{-},r^{-}).
\end{equation*}
Substituting it into the second equation, one obtains
\begin{equation*} 
\frac{\partial\chi}{\partial r^{-}} (q,q^{-},r^{-})
=
k_{2}(\varphi_{1}(q,q^{-})r^{-}+\tilde{\psi}_{1}(q,q^{-}))-\tilde{\psi}_{2}(q,q^{-}) ,
\end{equation*}
which gives
\begin{equation*} 
\chi   (q,q^{-},r^{-})
=
k_{2} \left(\frac{1}{2}\varphi_{1}(q,q^{-}) (r^{-}) ^{2}
+\tilde{\psi}_{1}(q,q^{-})r^{-} \right)
-\tilde{\psi}_{2}(q ,q^{-})r^{-}
- \tilde{\phi}(q,q^{-}).
\end{equation*}
Thus,
\begin{multline}   \label{eq:mar19.1-2}
(1-k_{1}k_{2})L
=\frac{1}{2}\varphi_{1}(q,q^{-})
\left(k_{1}r^{2}+2r^{-}r+k_{2}r^{-}\,^{2}\right)
\\
+\tilde{\psi}_{1}(q,q^{-})(r+k_{2}r^{-})-
\tilde{\psi}_{2}(q,q^{-})(k_{1}r+r^{-})
-\tilde{\phi}(q,q^{-}).
\end{multline}
In this case
\begin{equation*}
\alpha_{3}= k_{1} \alpha_{4} ,
\qquad
\alpha_{2}= k_{2} \alpha_{1} ,
\qquad
\Delta=( 1 - k_{1}k_{2} )\alpha_{1}\alpha_{4}.
\end{equation*}

Introducing the notations
\begin{equation*}
k_{1}=\frac{\alpha}{\beta},
\qquad
k_{2}=\frac{\gamma}{\beta},
\qquad
\varphi_{1}=    \frac{   \beta^{2}  -  \alpha \gamma }{\beta},
\end{equation*}
\begin{equation*}
\tilde{\psi}_{1}= - \frac{  \beta^{2} -  \alpha\gamma  }{\beta}   \psi_{1},
\qquad
\tilde{\psi}_{2}= - \frac{  \beta^{2} -  \alpha\gamma  }{\beta}    \psi_{2},
\qquad
\tilde{\phi}=  \frac{  \beta^{2} -  \alpha\gamma  }{\beta^{2}}\phi,
\end{equation*}
relation~(\ref{eq:mar19.1-2}) is simplified
\begin{equation}    \label{eq:mar21.10}
L
= { \alpha \over 2 }  r^{2} 
+ \beta r  r^{-}
+ { \gamma \over 2 }  ( r^{-} ) ^{2}
+ (\alpha\psi_{2}-\beta\psi_{1}) r 
+ (\beta\psi_{2}-\gamma\psi_{1})  r^{-}
- \phi.
\end{equation}
We also get 
\begin{equation*}
\alpha_{3}= \frac{\alpha}{\beta} \alpha_{4}  ,
\qquad
\alpha_{2}= \frac{\gamma}{\beta} \alpha_{1} ,
\qquad
\Delta =  \left( 1  -    \frac{\alpha\gamma}{\beta^2 } \right) 
\alpha_{1}\alpha_{4}. 
\end{equation*}

With Lagrangian  (\ref{eq:mar21.10}) and, therefore, 
\begin{equation*}
L^{+}
= { \alpha^{+}  \over 2 } ( r^{+}) ^{2}
+ \beta^{+}  r^{+} r 
+ { \gamma^{+} \over 2 } r ^{2}
+ (\alpha^{+}\psi_{2}^{+}-\beta^{+}\psi_{1}^{+}) r^{+}
+ (\beta^{+}\psi_{2}^{+}-\gamma^{+}\psi_{1}^{+}) r 
- \phi^{+}, 
\end{equation*}
we rewrite equation~(\ref{equt1-1}) as
\begin{equation*}
\left( 
\frac{\beta}{\alpha_{4}}
-\frac{\beta^{+}}{\alpha_{1}^{+}}
\right) r 
+\frac{\beta^{+}}{\alpha_{1}^{+}}\psi_{1}^{+}
+\frac{\beta}{\alpha_{4}}\psi_{2}
=0.
\end{equation*}
Splitting it with respect to $r$, one obtains
\begin{equation*}
\frac{\beta^{+}}{\alpha_{1}^{+}}
=\frac{\beta}{\alpha_{4}}, 
\qquad  
\psi_{1}^{+}=-\psi_{2}. 
\end{equation*}


Using differentiation of these equations with respect to $q^{-}$ and $q$, one gets that
\begin{equation*}
\alpha_{1}  ( q, q^- )   =  \mu (q ^- )  \beta   ( q, q^- )  , 
\qquad 
\alpha_{4}  ( q, q^- )   =  \mu (q )  \beta   ( q, q^- )  , 
\end{equation*}
\begin{equation*}
\psi_{1} ( q, q^- )   =  \lambda  (q ^-) , 
\qquad 
\psi_{2 }  ( q, q^- ) =    -  \lambda  (q ) . 
\end{equation*}
where  $\lambda (q)$ and  $\mu(q)$ are some functions. 
In this case equations~(\ref{eq:gen-1}) give
\begin{equation}
p=\frac{ r -  \lambda   (q) }{ \mu (q ) }, 
\qquad 
p^{-}= \frac{ r^{-} -  \lambda   (q^- )}{\mu (q^-) }. 
\end{equation}
Hence, solving the latter equations, one has
\begin{equation}   \label{eq:mar21.22}
r=   \mu (q) p  +  \lambda  (q), 
\qquad
r^{-}=  \mu (q^-  ) p^{-}  + \lambda   (q^- ). 
\end{equation}
Lagrangian~(\ref{eq:mar21.10})  takes the form
\begin{equation}   \label{eq:mar21.10_mod}
L 
= { \alpha \over 2 }   {\dot{q}}^{2} 
+ \beta \dot{q}\dot{q}^{-} 
+ { \gamma \over 2 } (\dot{q}^{-})^{2}
- ( \alpha \lambda    + \beta  \lambda ^-  )  \dot{q} 
- ( \beta   \lambda   + \gamma \lambda  ^-  )  \dot{q} ^- 
-  \phi. 
\end{equation}

Further, we show that the Lagrangian approach~\cite{bk:DorodnitsynKozlovMeleshko[2023]} and developed above Hamiltonian approach are equivalent. 
The Elsgolts equation for the Lagrangian~(\ref{eq:mar21.10_mod}) is
\begin{multline}  \label{eq:Elsgoltz}
-   \beta ^+  \ddot{q}^{+}
-   ( \alpha  + \gamma ^+ ) \ddot{q} 
-    \beta   \ddot{q}^{-} 
\\
+   \left(  { \alpha_{q}  ^+ \over 2 }  -   \beta^+ _{{q}^{+}} \right)  (\dot{q}^+) ^{2}
-  \gamma  ^+ _{{q}^{+}}  \dot{q}^+   \dot{q}
-     { \alpha  _q   +  \gamma ^+ _q  \over 2 }  \dot{q} ^2   
- \alpha   _{{q}^{-}}  \dot{q}   \dot{q} ^- 
+  \left(  { \gamma_{q}  \over 2 }   -   \beta  _{{q}^{-}} \right)  (\dot{q}^-) ^{2}
\\
+ (  \beta ^+   (  \dot{\lambda} ^+   -    \dot{\lambda}  ) 
+   (  \beta ^+   _{{q}^{+}}  -  \alpha ^+   _{q}   )    \lambda   ^+ 
+  (  \gamma ^+   _{{q}^{+}}  -  \beta  ^+   _{q}   )    \lambda   )     \dot{q}^+
\\
+ (  \beta   (   \dot{\lambda} ^-  -     \dot{\lambda}   ) 
+   (  \alpha   _{{q}^{-}}  -  \beta     _{q}   )    \lambda   
+  (  \beta     _{{q}^{-}}  -  \gamma      _{q}   )    \lambda  ^-   )     \dot{q}^- 
\\ 
-\phi_{q}-\phi_{q}^{+}
=0
\end{multline}

Applying the Legendre transformation    (\ref{eq:apr3.4})
with coefficients
\begin{equation} 
\alpha _1   = \beta  \mu ^- , 
\qquad
\alpha _2   = \gamma  \mu ^- , 
\qquad
\alpha _3   = \alpha  \mu  , 
\qquad
\alpha _4   = \beta  \mu  , 
\end{equation} 
we obtain the Hamiltonian function 
\begin{equation}    \label{Hamiltonian_ex}
H
=  { \alpha  \over 2 }   ( {  \mu p  +  \lambda   } ) ^{2} 
+  \beta   ( {  \mu p   +  \lambda } )   ( {  \mu  ^-  p ^-   + \lambda ^- } )
+ { \gamma  \over 2 }   ( {  \mu  ^-  p ^-   + \lambda ^- }  ) ^{2}
+  \phi. 
\end{equation}
Further, the delay canonical Hamiltonian equations are 
\begin{subequations}  
\begin{gather}  
\label{eq:el_pha_ex}
\mu ( \beta  ^+    \dot{q} ^+ 
+ (   \alpha  +   \gamma ^+ )  \dot{q}
+ \beta    \dot{q} ^-   ) 
=  
H_{p} +H_{p}^{+}, 
\\
\label{eq:el_qha_ex}
\begin{array}{c}
D ( 
\beta  ^+   \mu ^+   p ^+ 
+ (   \alpha   +   \gamma ^+  )  \mu  p
+ \beta   \mu ^-     p ^- 
) 
= 
-  H_{q}  -  H_{q}^{+}. 
\end{array}
\end{gather}
\end{subequations}

Substituting Hamiltonian   (\ref{Hamiltonian_ex}) into equation (\ref{eq:el_pha_ex}), we get
\begin{multline*} 
\mu ( \beta  ^+    \dot{q} ^+ 
+ (   \alpha  +   \gamma ^+ )  \dot{q}
+ \beta    \dot{q} ^-   )  
\\
=  
 \alpha  (  \mu  p    +  \lambda  )  \mu 
+  \beta  (  \mu ^-   p ^-   +  \lambda  ^-  )  \mu 
+ \beta ^+  (  \mu ^+   p ^+   +  \lambda  ^+  )  \mu 
+ \gamma ^+ (  \mu    p   +  \lambda   )  \mu 
\end{multline*} 
Hence, 
\begin{equation*}
\beta  ^+   \mu ^+   p ^+ 
+ (   \alpha   +   \gamma ^+  )  \mu  p
+ \beta  ^-  \mu ^-     p ^- 
= 
( \beta  ^+    \dot{q} ^+ 
+ (   \alpha  +   \gamma ^+ )  \dot{q}
+ \beta    \dot{q} ^-   ) 
- 
(   \alpha  \lambda 
+\beta  \lambda  ^- 
+ \beta ^+ \lambda  ^+ 
+ \gamma ^+   \lambda  ). 
\end{equation*}
We differentiate this equation and substitute its right-hand side 
for the left-hand side of equation   (\ref{eq:el_qha_ex}).  
Eliminating $p^+$,   $p$,   and $p^-$ with the help of 
\begin{equation*}
p ^+ =\frac{ \dot{q} ^+- \lambda  ^+ }{\mu  ^+ }, 
\qquad 
p=\frac{ \dot{q} - \lambda  }{\mu }, 
\qquad 
p^{-}=\frac{ \dot{q}^{-} - \lambda  ^- }{\mu^- },
\end{equation*}
we obtains the Elsgolts equation~(\ref{eq:Elsgoltz}). 
Thus, we showed that the Lagrangian and Hamiltonian approaches developed above are equivalent.

\begin{remark}
Further generalizations of the proposed above delay Legendre transformation (\ref{eq:apr3.4}) are possible. 
For example, one can consider time-dependent Lagrangians 
$ L ( t, t^- , q  , q^-   p, p ^- )$.    
In all cases, the idea of the Legendre transformation's self-consistency at the given and shifted points should work.
\end{remark}

\section{Alternative approach for deriving coefficients  $\alpha_{i}   $}
\label{Alternative_approach}

We present an alternative approach to obtain coefficients
$\alpha_{i}$ based on assumption that in a joint space $(q, \dot q,
p)$ the Legendre relation should link coordinates $p$ and $p^- $ with
$\dot{q}$ and $ \dot{q}^-$ by a  {\it point transformation}. In that case
coefficients~(\ref{coefficients}) can be obtained if we resolve
relations~(\ref{relations}) as
\begin{subequations}
\begin{gather}  \label{point_relation_1}
\alpha_{1}p^{-}
=
\beta \dot{q}^{-},
\\
\alpha_{3}p
=
\alpha  \dot{q},
\\
 \label{point_relation_3}
\alpha_{2}p^{-}
=
\gamma \dot{q}^{-},
\\
\alpha_{4}p
=
\beta \dot{q}.
\end{gather}
\end{subequations}
Applying   compatibility conditions  $ p = S_+ ( p^-) $ and  $ \dot{q} = S_+ ( \dot{q}^-)    $
to equations~(\ref{point_relation_1}) and~(\ref{point_relation_3}),
we obtain
\begin{subequations}
\begin{gather}
\alpha_{1}p
=
\beta \dot{q},
\\
\alpha_{3}p
=
\alpha  \dot{q},
\\
\alpha_{2}p
=
\gamma \dot{q},
\\
\alpha_{4}p
=
\beta \dot{q}.
\end{gather}
\end{subequations}
These relations lead to coefficients~(\ref{coefficients}). 
Note that this approach treats the  {non}degenerate and degenerate cases together.

\section{Delay ({non}degenerate) harmonic oscillator without connection to Lagrangian}
\label{H_without_L}

We consider the Hamiltonian from Example  \ref{Non_degenerate case} 
without any connection with the Lagrangian 
and examine a {non}consistent way to introduce  coefficients  $\alpha_{i}$.

\begin{example}

We consider  the Hamiltonian 
\begin{equation}
H=pp^{-}+qq^{-}
\end{equation} 
with  coefficients 
\begin{equation*}
\alpha_{1}=0, 
\qquad
\alpha_{2}=0,
\qquad
\alpha_{3}=1,
\qquad
\alpha_{4}=0
\end{equation*}
for the Hamiltonian elementary action in functional (\ref{functional_H_delay}).


Then,  we get the delay Hamiltonian equations 
\begin{subequations}   \label{HamEq321}
\begin{gather}
\dot{q}
=p^{+}+p^{-},
\\
\dot{p}
=- q^{+} -  q^{-}.
\end{gather}
\end{subequations}
Notice that two delay equations~(\ref{HamEq321})  are equivalent to one second-order equation with four delays.


 The system of delay equations~(\ref{HamEq321}) admits
 the following symmetry operators:
\begin{equation}
X_{1}={\frac{\partial}{\partial t}},
\qquad
X_{2}=q{\frac{\partial}{\partial q}}+p{\frac{\partial}{\partial p}},
\qquad 
X_{3}=p{\frac{\partial}{\partial q}}-q{\frac{\partial}{\partial p}}. 
\end{equation}
However, for variational symmetry  $ X_{1} $, we need to consider other equations than Eqs.  (\ref{HamEq321}). 
Symmetries  $ X_{2}$  and $ X_{3}$  are neither variational nor divergence symmetries of the Hamiltonian function: 
\begin{equation*}
\Omega _2 
=  2 \tilde{H}
= 2  ( p \dot{q} - H ) , 
\qquad
\Omega _3 
= q\dot{q} + p\dot{p} -2pq^- - 2p^-q.
\end{equation*}
Thus, these symmetries do not provide conserved quantities for equations    (\ref{HamEq321}).

One can search coefficients $\alpha_i$,  $ i = 1,2,3,4$,  for which the Hamiltonian is invariant
for the symmetries 
\begin{equation}
Y_{1}=\sin t {\frac{\partial}{\partial q}} + \cos t  {\frac{\partial}{\partial p}},
\qquad
Y_{2}=\cos t {\frac{\partial}{\partial q}} - \sin t {\frac{\partial}{\partial p}},
\end{equation}
admitted by a harmonic oscillator without delay. 
The  Hamiltonian is divergence invariant for the coefficients 
\begin{equation*}
 \alpha_{1}=\alpha_{4}=1,
\qquad
\alpha_{2}=\alpha_{3}=0, 
\end{equation*}
which were considered in Example  \ref{Non_degenerate case}.


\end{example}


\begin{thebibliography}{10}





\bibitem{bk:Lie[1888]}
S.~Lie.
\newblock {K}lassifikation und {I}ntegration von gew\"ohnlichen
  {D}ifferentialgleichungen zwischen $x,y$,
die eine {G}ruppe von  {T}ransformationen gestatten {I, II}.
{\em Math. Ann.}  {\bf 32}  213--281, 1888.
Gesammelte Abhandlungen, vol. 5, B.G. Teubner, Leipzig,
1924, pp.   240--310.

\bibitem{bk:Lie1924}
S.~Lie,
Gruppenregister,
{\em Gesammelte Abhandlungen},
{\bf 5} 767--773, 1924.

\bibitem{bk:Ovsiannikov1978}
L.~V.~Ovsiannikov,
{\em Group Analysis of Differential Equations},
Nauka, Moscow,
1978.
{E}nglish translation:  {W}.~{F.}~{A}mes, Ed.,
Academic  Press, New York, 1982.

\bibitem{bk:Ibragimov[1983]}
N.~H.~Ibragimov,
{\em Transformation Groups Applied to Mathematical Physics},
Nauka, Moscow,
1983.
{E}nglish translation:  Reidel, D., Ed., Dordrecht, 1985.

\bibitem{bk:Olver[1986]}
P.~J.~Olver,
{\em Applications of {L}ie Groups to Differential Equations},
Springer-Verlag, New York,
1986.

\bibitem{bk:Gaeta1994}
G.~Gaeta,
{\em Nonlinear Symmetries and Nonlinear Equations},
Kluwer, Dordrecht,
1994.

\bibitem{bk:HandbookLie}
N.~H.~Ibragimov, Ed.,
{\em {CRC} Handbook of {L}ie Group Analysis of Differential Equations},
volume 1, 2, 3,
CRC Press, Boca Raton,
1994, 1995, 1996.

\bibitem{bk:BlumanAnco2002}
G.~W.~Bluman and S.~C.~Anco,
{\em Symmetry and Integration Methods for Differential Equations},
Springer, New York, 2002.

\bibitem{Noether1918}
E.~Noether,
Invariante variations problem,
{\em Nachr. d. {K}\"oniglichen Gesellschaft der Wissenschaften zu
  G\"ottingen, Nachrichten, Mathematisch-Physikalische Klasse Heft 2},
 pages 235--257,
1918.
English translation:
Transport Theory and Statist. Phys.,
{\bf 1} (3) 183-207,
1971,
(arXiv:physics/0503066[physics.hist-ph]).

\bibitem{bk:AncoBluman1997}
S.~C.~Anco and G.~Bluman,
Direct construction of conservation laws from field equations,
{\em Phys. Rev. Lett.} 
{\bf 78}(3) 2869--2873,
1997.









\bibitem{Dorodnitsyn1991}
V.~A. Dorodnitsyn,
Transformation groups in net spaces,
{\em Journal of Soviet Mathematics},
{\bf 55}(1)  1490--1517,
1991.

\bibitem{LeviWinternitz1991}
D.~Levi and P.~Winternitz,
Continuous symmetries of discrete equations,
{\it Phys. Lett.} {\bf  152} 335--338,
1991.

\bibitem{QuispelCapelSahadevan}
G.~R.~W.~Quispel, H.~W.~Capel  and R.~Sahadevan,
Continuous symmetries of differential--difference equations,
{\em Phys. Lett. A},
{\bf 170}(5)  379--383,
1992.

\bibitem{Dorodnitsyn1993a}
V.~A.~Dorodnitsyn,
The finite-difference analogy of Noether's theorem,
{\em Doklady RAN}, {\bf 328}(6) 678--682,
1993. 
English translation: {\it Phys. Dokl.} {\bf 38}(2) 66--68.

\bibitem{Dorodnitsyn1993}
V.~A.~Dorodnitsyn,
Finite-difference models entirely inheriting symmetry of original differential equations,
In {\em Modern Group Analysis: Advanced Analytical and Computational  Methods in Mathematical Physics},
volume 191. Kluwer Academic Publishers, Boston,
1993.

\bibitem{DorodnitsynKozlovWinternitz2000}
V.~Dorodnitsyn, R.~Kozlov and P.~Winternitz,
Lie group classification of second-order ordinary difference equations,
{\em J. Math. Phys.},
{\bf 41}(1)  480--504,
2000.

\bibitem{DorKozWin2004}
V.~Dorodnitsyn, R.~Kozlov, and P.~Winternitz,
Continuous symmetries of Lagrangians and exact solutions of discrete equations,
{\em J. Math. Phys.},
{\bf 45}(1) 336--359,
2004.

\bibitem{LeviWinternitz2005}
D.~Levi and P.~Winternitz,
Continuous symmetries of difference equations,
{\em J. Phys. A: Math. Gen.},
{\bf 39}(2)  R1--R63,
2006.

\bibitem{bk:Dorodnitsyn[2011]}
V.~A.~Dorodnitsyn,
{\em Applications of Lie Groups to Difference Equations},
CRC Press, Boca Raton,
2011.

\bibitem{Winternitz2011}
P.~Winternitz,
Symmetry preserving discretization of differential equations
and Lie point symmetries of differential--difference equations,
in  D.~Levi  {\it et al}  Eds.
{\it Symmetries and Integrability of Difference Equations},
pp. 292--341,
Cambridge, Cambridge University Press,
2011.
London Mathematical Society Lecture Notes.

\bibitem{bk:Hydon2014}
P.~E.~Hydon,
{\em Difference Equations by Differential Equation Methods},
Cambridge University Press, Cambridge,
2014.

\bibitem{bk:DKapKozWin}
V.~A.~Dorodnitsyn, E.~I.~Kaptsov, R.~V.~Kozlov and P.~Winternitz,
The adjoint equation method for constructing first integrals of difference  equations,
{\it J. Phys. A: Math. Theor.},
{\bf 48}(5) 055202,
2015.

\bibitem{bk:Zhang}
Y.~Zhang,
Noether symmetry and conserved quantity for a time-delayed Hamiltonian system of Herglotz type,
{\it R. Soc. Open Sci.}
 {\bf 5}(10), 180208,
2018.











\bibitem{bk:ChevDK}
A.~F.~Cheviakov, V.~A.~Dorodnitsyn and E.~I.~Kaptsov,
Invariant conservation law-preserving discretizations of linear and nonlinear wave equations,
 {\it J. Math. Phys.},
{\bf 61}, 081504,
2020.

\bibitem{bk:DKap1}
V.~A.~Dorodnitsyn and E.~I.~Kaptsov,
Shallow water equations in Lagrangian coordinates:
Symmetries, conservation laws and its preservation in difference models,
{\it Commun. Nonlinear Sci. Numer. Simulat.}
{\bf  89},   105343,
2020.

\bibitem{bk:DKap2}
V.~ A.~Dorodnitsyn and E.~I.~Kaptsov,
Discrete shallow water equations preserving symmetries and conservation laws,
{\it J. Math. Phys.}
{\bf 62}(8)  083508,
2021.

\bibitem{bk:DKapMel}
V.~A.~Dorodnitsyn, E.~I.~Kaptsov, and S.~V.~Meleshko,
Symmetries, conservation laws, invariant solutions and difference schemes of the one-dimensional Green-Naghdi equations, {\it J. Nonl. Math.  Phys.},
{\bf  28}(1)  90-107,
2021.








\bibitem{bk:Elsgolts[1955]}
L.~E.~El'sgol'c,
{\it Qualitative Methods in Mathematical Analysis},
GITTL, Moscow,
1955.
English translation:
American Mathematical Society (Translations of Mathematical Monographs), 1964.

\bibitem{Driver1977}
R.~D.~Driver,
{\it Ordinary and delay differential equations},
Springer-Verlag,  New York,
1977.

\bibitem{Erneux}
T.~Erneux,
{\it Applied Delay Differential Equations},
Springer,
2009.

\bibitem{Polyanin2023}
A.~D.~Polyanin, V.~G.~Sorokin, and A.~I.~Zhurov,
{\it Delay Ordinary and Partial Differential Equations (Advances in Applied Mathematics)},
Chapman and Hall/CRC,
2023.













\bibitem{bk:DorodnitsynKozlovMeleshkoWinternitz[2018a]}
V.~A.~Dorodnitsyn, R.~Kozlov, S.~V.~Meleshko and P.~Winternitz,
Lie group classification of first-order delay ordinary differential equations,
{\it J. Phys. A: Math. Theor.}
{\bf 51} 205202,
2018.

\bibitem{bk:DorodnitsynKozlovMeleshkoWinternitz[2018b]}
V.~A.~Dorodnitsyn, R.~Kozlov, S.~V.~Meleshko and P.~Winternitz,
Linear or linearizable first-order delay ordinary differential equations and their lie point symmetries,
{\it J. Phys. A: Math. Theor.}
{\bf 51}  205203,
2018.

\bibitem{bk:DorodnitsynKozlovMeleshkoWinternitz2021}
V.~A.~Dorodnitsyn, R.~Kozlov, S.~V.~Meleshko and P.~Winternitz,
Second-order delay ordinary differential equations, their symmetries and application to a traffic problem,
{\it J. Phys. A: Math. Theor.}
{\bf 54}   105204,
2021.


\bibitem{bk:DorodnitsynKozlovMeleshko[2023]}
V.Dorodnitsyn, R.Kozlov, S. Meleshko,
Lagrangian formalism and Noether-type theorems for second-order delay ordinary differential equations,
{\it J. Phys. A: Math. Theor.}
{\bf 56}(34)  345203,
2023.








\bibitem{Gelfand}
I.~M.~Gelfand and S.~V.~Fomin,
{\it  Calculus of variations},
Prentice--Hall, Inc., Englewood Cliffs, NJ,
1963.

\bibitem{Abraham}
R.~Abraham and J.~E.~Marsden,
{\it Foundations of mechanics},
Benjamin/Cummings Publishing Co., Inc., Reading, MA,
1978.

\bibitem{Goldstein}
H.~Goldstein,
{\it Classical mechanics},
Addison-Wesley Publishing Co., Reading, MA,
1980.

\bibitem{Arnold} V.~I.~Arnold,
{\it Mathematical methods of classical mechanics},
Springer--Verlag, New York,
1989.

\bibitem{Marsden}
J.~E.~Marsden and T.~S.~Ratiu,
{\it Introduction to mechanics and symmetry. A basic exposition of classical mechanical systems},
Springer--Verlag, New York,
1999.

\bibitem{bk:DorodnitsynKozlov}
V.~Dorodnitsyn and R.~Kozlov,
{L}agrangian and {H}amiltonian formalism for discrete equations:   Symmetries and first integrals,
in D.~Levi  {\it et al}  Eds.
{\em Symmetries and Integrability of Difference Equations},
pages 7--49,  Cambridge University Press, Cambridge,
2011.
London Mathematical Society Lecture Notes.

\bibitem{bk:DorodnitsynKozlov2009}
V.~Dorodnitsyn and R.~Kozlov,
First integrals of difference Hamiltonian equations,
{\it J. Phys. A: Math. Theor.} 
{\bf 42}(45) 454007,
2009.

\bibitem{bk:DorodnitsynKozlov2010}
V.~Dorodnitsyn and R.~Kozlov,
Invariance and first integrals of continuous and discrete Hamiltonian equations,
{\it Journal of Engineering Mathematics} 
{\bf 66}(1) 253--270,
2010.

\bibitem{Elsgolts2}
L.~E.~Elsgolts,
Variational problems with retarded arguments.
{\em Vestnik Moskovskogo Universiteta. Seriya 1.
Matematika.   Mekhanika.}
{\bf 10} 57--62,
1952.
(in Russian)

\bibitem{Elsgolts3}
L.~E.~Elsgolts,
Variational problems with retarded arguments,
{\em Uspekhi Matematicheskikh Nauk} 
{\bf 12} (1(73)) 257--258,
1957.
(in Russian)

\bibitem{Bess}
E.~Bessel-Hagen,
 \"{U}ber die {E}rhaltungssatze der {E}lektrodynamik,
{\em Math. Ann.} 
{\bf 84} 258--276,
1921.

\bibitem{bk:Ostrogradsky}
M.~V. Ostrogradsky,
Memoires sur les \'{e}quations diff\'{e}rentielles, relatives au  probl\`{e}me des isop\'{e}rim\`{e}tres,
{\em M. Mem. Acad. St.Petersbourg} 
{\bf 6}(4) 385--517,
1850.




\end{thebibliography}
\end{document}